\documentclass[letterpaper, 10 pt, journal, twoside]{IEEEtran}

\usepackage[%
    style=ieee,
    sorting=none,
    natbib=true,  
    backend=biber,
    sortcites=true,
    doi=false,
    url=false,
    mincitenames=1,
    maxcitenames=1, 
    minbibnames=2, 
    maxbibnames=3, 
    hyperref
]{biblatex}

\usepackage[nolist,printonlyused]{acronym}
\usepackage{amsmath, amssymb, amsfonts}

\allowdisplaybreaks
\usepackage{amsthm}
\usepackage{algorithm}
\usepackage{algpseudocode}
\usepackage{balance}
\usepackage{bbold}
\usepackage{bm}
\usepackage[english]{datetime2}
\usepackage{glossaries}
\usepackage{multicol}
\usepackage{textcomp}

\usepackage{sidecap}
\sidecaptionvpos{figure}{t}

\usepackage[parse-numbers=false]{siunitx}
\usepackage{xcolor}

\makeatletter
\let\MYcaption\@makecaption
\makeatother

\usepackage[font=footnotesize]{subcaption}

\makeatletter
\let\@makecaption\MYcaption
\makeatother

\usepackage{graphicx}
\usepackage{hyperref}
\usepackage{cleveref}
\usepackage{mathtools}

\newtheorem{assumption}{Assumption}

\usepackage[switch]{lineno}

\newtheorem{remark}{Remark}

\crefname{figure}{Fig.}{Figs.}
\crefname{equation}{}{}

\crefname{table}{Table}{Tables}
\crefname{algorithm}{Alg.}{Algorithms}
\crefname{section}{Section}{Sections}
\crefname{remark}{Remark}{Remarks}
\def\BibTeX{{\rm B\kern-.05em{\sc i\kern-.025em b}\kern-.08em
    T\kern-.1667em\lower.7ex\hbox{E}\kern-.125emX}}

\newcommand\hmzh[1]{\textcolor{purple}{[HK: #1]}}

\newcommand\topic[1]{\textcolor{purple}{#1}}

\renewcommand\hmzh[1]{#1}
\renewcommand\topic[1]{#1}

\newcommand\subdataset[1]{\dataset{}_{#1}}
\newcommand\wptset[1]{N_{#1}}
\newcommand\traj[1]{\xi^{#1}}
\newcommand\autotraj[1]{{\xi_a}^{#1}}
\newcommand\trajspace{\Xi}

\newcommand\state[1]{\textbf{x}_{#1}}
\newcommand\stategt[2]{\textbf{x}^{\text{GT}, #2}_{#1}}
\newcommand\ctrl[1]{\textbf{u}_{#1}}

\newcommand\numstates{n}
\newcommand\numctrls{r}
\newcommand\numdatapts{m}

\newcommand\horizon{T}
\newcommand\natset[2]{\mathcal{N}^{#1}_{#2}}

\newcommand\dynamics[1]{f_{#1}}

\newcommand\kmeans{$k$-means}

\newcommand\nstates[1]{\mathcal{X}^{#1}}

\newcommand{\numconvexsets}[1]{k_{#1}}
\newcommand{\convsubset}[2]{C_{#1}^{#2}}

\newcommand\dataset{\mathcal{D}}

\newcommand\numhullstates{n_c}
\newcommand\hullstate[1]{y_{#1}}

\newcommand\nset{naturalistic behavior set}
\newcommand\nsets{naturalistic behavior sets}
\newcommand\Nset{Naturalistic behavior set}
\newcommand\NSET{Naturalistic Behavior Set}

\begin{document}

\begin{acronym}
    \acro{RV}{random variable}
    \acro{ILQR}{Iterative Linear Quadratic Regulation}
    \acro{ILQGames}{Iterative Linear-Quadratic Games}
    \acro{SILQGames}{Stackelberg Iterative Linear-Quadratic Games}
    \acro{SLF}{Stackelberg Leadership Filter}
    \acro{LQ}{linear-quadratic}
\end{acronym}

\title{Act~Natural!~Extending~Naturalistic~Projection to~Multimodal~Behavior~Scenarios}

\author{Hamzah I.\ Khan and David Fridovich-Keil\thanks{
This work was supported by the National Science Foundation under Grants 2211548 and 2336840.
The authors ({\tt\small \{hamzah, dfk\}@utexas.edu}) are with the \textit{Department of Aerospace Engineering and Engineering Mechanics}, University of Texas at Austin.}
}


\maketitle

\begin{abstract}
Autonomous agents operating in public spaces 
must consider how their behaviors might affect the humans around them, 
even when not directly interacting with them.
To this end, it is often beneficial
to be predictable and appear naturalistic.
Existing methods for this purpose 
use human actor intent modeling or imitation learning techniques, but these approaches rarely capture all possible motivations for human behavior and/or require significant amounts of data.
Our work extends a technique for modeling unimodal naturalistic behaviors with an explicit convex set representation,
to account for multimodal behavior by using multiple convex sets.
This more flexible representation provides a higher degree of fidelity in data-driven modeling of naturalistic behavior that arises in real-world scenarios in which human behavior is, in some sense, discrete, e.g. whether or not to yield at a roundabout.
Equipped with this new set representation, we develop an
optimization-based filter to project
arbitrary trajectories into the set so that they appear naturalistic to humans in the scene, while also satisfying vehicle dynamics, actuator limits, etc.
We demonstrate our methods on real-world human driving data from the inD (intersection) and rounD (roundabout) datasets.


\end{abstract}

\section{Introduction}
\label{sec:introduction}

Safe and comfortable interaction between humans and autonomous agents requires a measure of predictable and naturalistic behavior from autonomous systems.
Autonomous agents in the real world can easily find themselves in unsafe situations when they violate these informal norms of human-like behavior.
As one example, autonomous vehicles are well-documented to behave more cautiously than human drivers expect, which can lead to
human drivers reacting unsafely to unexpected or abnormal driving and causing collisions \citep{teoh2017google}.
The requirement to act like other agents is even written into some traffic laws,\footnote{Texas Transportation Code §545.363} in which vehicles on the road must follow the ``flow-of-traffic,'' i.e. the speed at which other vehicles are moving.
Thus, autonomous agents must be able to plan and execute naturalistic, human-like behavior.

However, naturalistic behavior tends to be challenging to model mathematically because human preferences and decision-making are opaque.
Nevertheless, there exists a need for techniques which are able to model the wide variety of naturalistic behavior based on observations of human actions.
Prior works in this area tend to fall short in modeling the diversity of human behavior \citep{bajcsy2021analyzing,sadigh2016infogatheringonhumans}, or require significant amounts of data which may not always be available \citep{kuefler2017imitating,bhattacharyya2023generative}, or are limited in the types of scenarios they can model \citep{khan2024actnatural}.
These deficiencies present a need 
for improved modeling of naturalistic behavior for use in downstream planning tasks.

\begin{figure}[!t]
    \centering
    \includegraphics[width=\columnwidth]{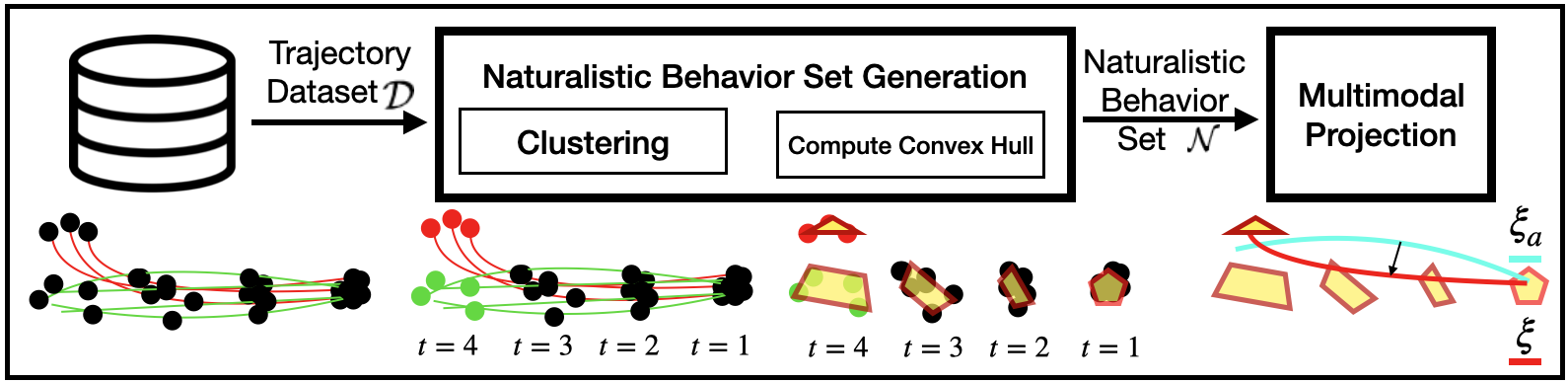}
    \vspace{-12pt}
    \includegraphics[width=\columnwidth]{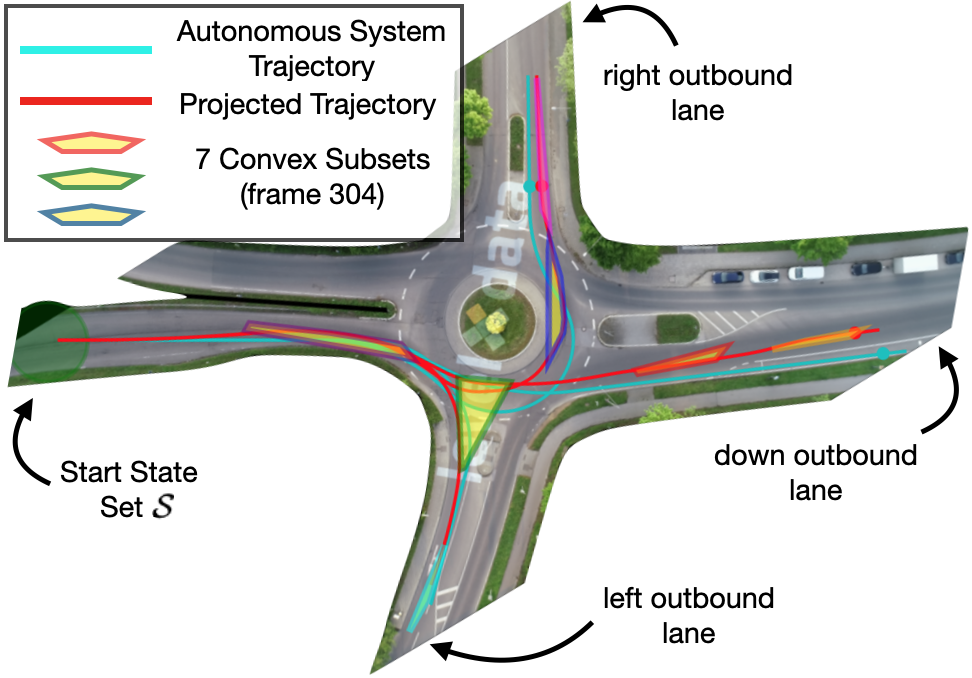}
    \caption{\label{fig:rounD-r0-front-figure-7means-f304}
    \textbf{(top)} Given a multimodal behavior dataset $\dataset{}$, our method first generates a \nset{} by computing nonconvex time-indexed sets.
    Each nonconvex set is represented by the union of a set of convex hulls which are formed by clustering the trajectory states at each time.
    Then, we project arbitrary trajectories into this set to make the behaviors more naturalistic.
    \textbf{(bottom)} Demonstration of our method on a roundabout scenario. Observe that the polygons cover the entry and exit lanes, as well as regions around the roundabout.
    Subsequently, the projected trajectories (shown in red) more closely align with human data. Figure is rotated for presentation.
    \vspace{-1.5em}
    }
\end{figure}

In this work, we extend the naturalistic projection technique first introduced by \citet{khan2024actnatural}.
Specifically, we present a data-driven algorithm for representing the naturalistic set of human behaviors based on observation data of human behavior.
Our method utilizes convex hulls as a simple, interpretable, and computationally efficient way to form such a set, and we propose a projection method for making arbitrary autonomously planned trajectories naturalistic using mixed-integer optimization.
\citet{khan2024actnatural} notes a number of limitations including a need to define a ``single-task'' dataset and a lack of support for multimodal behaviors.
In this work, we loosen the requirements on the dataset from which we generate a \nset{} by redefining the representation of \nset{}s to model multimodal behavior such as driving through an intersection, which may have multiple exit points from any given entrance.
Sprecifically, we represent naturalistic behavior through the use of multiple convex hulls at each moment in time.
Thus, in contrast with \cite{khan2024actnatural}, our method enables a \nset{} to model vehicle behavior around a roundabout with multiple possible exits, as in \cref{fig:rounD-r0-front-figure-7means-f304}.

In this work, we develop a simple and efficient method for representing a variety of naturalistic behaviors, and a method for making autonomous system trajectories naturalistic by projecting them into our representation in a dynamically consistent manner.
We make four primary contributions:
\begin{enumerate}
    \item we extend the definition of \nset{}s proposed by \cite{khan2024actnatural} to represent multimodal behaviors as unions of convex sets at each time,
    \item we propose an updated method for generating this modified \nset{} which clusters behavior data at each time and then constructs convex hulls around each cluster of data,
    \item we propose a mixed-integer optimization-based projection method, adapted from the one proposed by \cite{khan2024actnatural}, that enforces naturalistic behavior on an arbitrary autonomous system trajectory, and
    \item we demonstrate the ability of \nset{} generation to capture patterns, which may not be explicitly modeled, in real-world human behavior data on the {inD} \cite{inD2020naturalisticdataset} and {rounD} \cite{rounDdataset} scenario driving datasets.
\end{enumerate}

\section{Related Work}
\label{sec:related-work}

Prior works \cite{dragan2013legibility,dragan2013legiblemotion} have observed the phenomenon of unexpected autonomous behavior leading to unsafe behavior caused by humans, and suggest that autonomous planners should anticipate how human actors respond to their behavior \citep{sadigh2016planning}.
\citet{dragan2013legibility},~\cite{dragan2013legiblemotion}
propose the idea of legible motion, where an observer can easily understand the intent or strategy through the actor's movements.
This is different from predictability, which involves easily inferred movements without necessarily understanding the intent.
\citet{sadigh2016planning} observe that the actions of an autonomous vehicle can unintentionally influence human drivers, which implies that autonomous vehicles need to model human behavior in their prediction and planning processes, as otherwise safe but unexpected actions might cause unsafe reactions from humans.

Existing methods have approached the problem of accounting for human behavior in one of two ways: one of the categories of methods infers intent from human behavior models.
The other involves imitating and reproducing human behavior based on observed data.
Human behavior models use various approaches. 
Some explicitly model and infer specific aspects of human internal states like rationality \citep{bobu2018learning,fridovich2019confidenceaware} and target states \citep{sadigh2016infogatheringonhumans}. 
Other models use data-driven predictive human behavior modeling to infer human actors' beliefs about other agents' goals \citep{bajcsy2021analyzing,peters2023online} or to influence human actors' (possibly wrong) internal models of autonomous agents' motion \citep{tian2023dynamicsofhumanlearning}.
However, these approaches often fall short in complex scenarios for which human preferences and behavior are naturally opaque and difficult to model mathematically.
In these situations, multiple unmodeled aspects of the world  may be significant, such as how road conditions may influence safety and comfort tolerances. Our approach, by contrast, frames naturalistic behavior planning as a projection into naturalistic constraints from human behavior data, without inferring the internal state of human actors.

The second set of approaches to this problem fall under the category of imitation learning, involving motion planning by mimicking provided human behavior data (i.e., behavior cloning and inverse reinforcement learning) \cite{bhattacharyya2023generative}.
Behavior cloning methods train supervised models on human data but tend to suffer from compounding errors and distribution shift \citep{bagnell2015invitation,kuefler2017imitating,DixonIRL2020}.
Inverse reinforcement learning techniques form a subset of imitation learning and infer a cost function describing an agent's objectives from data \citep{DixonIRL2020}.
These methods generalize better than behavior cloning and produce naturalistic human behavior appropriate to the specific setting in which an autonomous system operates.
However, it requires significantly more data \citep{kuefler2017imitating} and most inverse reinforcement learning approaches assume specific reward structures \citep{song2018imitationmultiagent,huang2021driving}.
For abnormal scenarios with limited data (i.e. weather-related, accident diversions, etc.), even imitation learning techniques are infeasible for modeling human behavior due to insufficient data.

Another method in this second category of accounting for human behavior is (unimodal) naturalistic projection \cite{khan2024actnatural}, which
replicates human-like behavior through the use of naturalistic data.
Specifically, \citet{khan2024actnatural} propose \nsets{}, a simple, general, and computationally efficient representation constructed from time-indexed convex hulls.
Using this representation, \cite{khan2024actnatural} projects existing autonomous system trajectories into the \nset{} by solving a convex optimization problem.
This approach produces intuitive results on some simple driving examples for which naturalistic projection produces intuitive driving behavior. 
However, it requires defining a single dataset of trajectories which represent similar behaviors.
This pre-filtering step makes the method less versatile
due to the practical challenge of heuristically filtering real data.
Additionally, the method performs poorly when trajectories represent similar behavior for parts of the trajectories, such as two trajectories driving into an intersection where one proceeds straight and the other turns right.
This particular example requires supporting multimodal naturalistic behaviors, which is not addressed by \cite{khan2024actnatural}.
Our approach in this paper aims to address precisely this limitation.

\begin{remark}[Inspirations from Approaches to Forward Reachability]
\label{rmk:forward-reachability}
Forward reachability analysis involves a set of techniques for modeling sets of feasible trajectories emanating from a given (set of) initial state(s) \citep{bansal2017reachability}.
While the forward reachability problem is fundamentally different than our own, both involve identifying sets that describe where agents may be at given times in the future.
Traditional approaches to forward reachability designed for uncertainty propagation provide robust over-approximations but tend to be excessively conservative, particularly for longer time horizons, thereby limiting their utility \citep{rober2023reachability}.
Techniques based on bounded intervals face similar limitations \citep{ramdani2011reachable}.
In contrast, our method offers a data-driven approximation of the multimodal naturalistic set, though it is inspired in part by some sample-based forward reachability frameworks (e.g., \citealp{lew2022simple, thorpe2022data}) which achieve improved results using simple data-driven techniques like generating convex hulls.
\end{remark}

\section{Problem Formulation}
\label{sec:formulation}
We consider a discrete-time dynamical system evolving over horizon $\horizon{} \in \mathbb{N}$, 
Let this system have state $\state{t} \in \nstates{} \subseteq \mathbb{R}^{\numstates{}}$ and control input $\ctrl{t} \in \mathcal{U} \subseteq \mathbb{R}^{\numctrls{}}$ defined at times $t \in [\horizon{}+1] \equiv \{ 0, 1, \ldots, \horizon{} \}$.
The system evolves 
according to dynamics
\begin{equation}
\state{t+1} = \dynamics{}(\state{t}, \ctrl{t}),
\label{eq:dynamics}
\end{equation}
starting from an initial state, $\state{0}$.
Next, we list our assumptions.
\begin{assumption}
\label{ass:known-dynamics}
We presume that dynamics $\dynamics{}$ in \eqref{eq:dynamics} are known.
\end{assumption}
\begin{assumption}
\label{ass:historical-data}
We assume access to historical trajectory data that describes ``well-behaved'' human behavior.
Each state trajectory is assumed to take the form
\begin{equation}
\label{eq:trajectory-form}
\traj{i} = [ \state{0}^{i\intercal} ~ \state{1}^{i\intercal} ~ \cdots ~ \state{\horizon}^{i\intercal} ]^\intercal \in \trajspace{} \subseteq \mathbb{R}^{\numstates{}(\horizon{} + 1)}.
\end{equation}
\end{assumption}
\noindent Well-behaved implies, for example, that these data should 
be considered safe, as humans generally avoid unsafe behavior.

\subsection{Estimating the \NSET{}} \label{ssec:formulation-estimating-nset}

Our objective is to derive a representation for the set of state trajectories that encapsulates the naturalistic behavior exhibited by a human operator while performing a task.
Such a task may involve, for instance, a driver beginning at a specified point passing through an intersection.
In such an example, we note that drivers may exit the intersection from different locations.

Formally, consider a dataset $\dataset{}$ comprising $m$ trajectories defined in \eqref{eq:trajectory-form}, corresponding to instances of a task:
\begin{equation}
\label{eq:dataset}
\dataset{} = \{ \traj{i} \}_{i=0}^{\numdatapts{}-1}.
\end{equation}
The goal is to estimate the \nset{} $\natset{}{}$, which represents the naturalistic behavior observed in the dataset.
Let $\natset{}{}$ be a time-indexed sequence of 
nonconvex subsets:
\begin{equation}
\label{eq:full-nat-set}
\natset{}{} = \{ \wptset{0}, \wptset{1}, \ldots, \wptset{\horizon} \} ,
\end{equation}
where each of the nonconvex subsets $\wptset{t}$ is represented by $\numconvexsets{t} \in \mathbb{N} \setminus \{ 0 \}$ convex subsets $\{ \convsubset{t}{i} \}$ of the state space $\mathcal{X}$,
\begin{equation}
\label{eq:nat-subset-at-t}
\wptset{t} = \bigcup_{i\in [k_t]} \convsubset{t}{i}, \qquad  \convsubset{t}{i} \subseteq \mathcal{X}.
\end{equation}
Thus, to estimate $\natset{}{}$, we solve two subproblems simultaneously using dataset $\dataset{}$. 
First, we must identify the number of convex subsets, $k_t$, needed to represent the nonconvex subset $\wptset{t}$ at each time.
Second, we must identify these convex subsets, $\{\convsubset{t}{i}\}$, at each time.

\begin{remark}[Difference from \cite{khan2024actnatural}] \citet{khan2024actnatural} define a \nset{} as a tube of time-indexed convex sets.
By contrast, our work defines a \nset{} as a set of time-indexed nonconvex subsets, where each time-indexed nonconvex subset is represented as a union of multiple convex sets.
Tuple $(i_0, i_1, \ldots, i_{\horizon})$ is an element of a set which indexes naturalistic tubes within $\natset{}{}$, where each naturalistic tube is a Cartesian product of convex sets $\{\convsubset{0}{i_0}, \convsubset{1}{i_1}, \ldots, \convsubset{\horizon{}}{i_{\horizon{}}}\} \in \natset{}{}$.
Hence, the definition of \nsets{} in \cite{khan2024actnatural} is a special case of our definition where $k_t=1$, i.e. there is a unique naturalistic tube in $\natset{}{}$ with index tuple $(0, 0, \ldots, 0)$.
\end{remark}
\noindent In both our work and \cite{khan2024actnatural}, an advantage of set-based representation is that they do not assume a particular distribution of the underlying data, which is critical to capturing the wide variety of human behavior.

\subsection{Projection Into the \NSET{}}
Because autonomous system trajectories do not demonstrate naturalistic behavior, the resulting behaviors may be unpredictable, and therefore unsafe, when operating around humans \citep{teoh2017google}.
Thus, we seek to augment an autonomously planned trajectory to more closely resemble natural behavior. 
Specifically, given such a trajectory, $\autotraj{}$, we seek to project $\autotraj{}$ into the learned \nset{} $\natset{}{}$.
This projected trajectory, $\xi = [\state{0}^\intercal ~ \state{1}^\intercal ~ \cdots ~ \state{\horizon{}}^\intercal]$, should also satisfy dynamic feasibility with respect to \eqref{eq:dynamics}.

We say a state $\state{t}$ is within naturalistic subset $\wptset{t}$ if 
$\state{t}$ is contained within at least one of the $\numconvexsets{t}$ convex sets in $\wptset{t}$, i.e.
\begin{equation}
\label{eq:naturalistic-constraint-at-t}
\state{t} \in \wptset{t} \equiv \left( \state{t} \in \convsubset{t}{0} \right) \vee \left( \state{t} \in \convsubset{t}{1} \right) \vee \cdots \vee \left( \state{t} \in \convsubset{t}{\numconvexsets{t}-1} \right).
\end{equation}
Building on this definition, we say that the trajectory $\xi$ is naturalistic if \cref{eq:naturalistic-constraint-at-t} at every time,
\begin{equation}
\label{eq:naturalistic-constraint}
\xi \in \natset{}{} \quad \equiv \quad \state{t} \in \wptset{t}, \forall t \in [\horizon{}+1].
\end{equation}

As with naturalistic set generation, the multimodal projection problem requires solving two simultaneous subproblems.
First, we must identify which naturalistic tube $(i_0, i_1, \ldots, i_{\horizon{}})$ to project $\autotraj{}$ into.
Second, we must perform the projection and solve for a dynamically feasible trajectory $\traj{}$ so that
\begin{equation}
\label{eq:naturalistic-trajectory-implication}
\xi \in \convsubset{0}{i_0} \times \convsubset{1}{i_1} \times \cdots \times \convsubset{\horizon{}}{i_\horizon{}}
\end{equation}
and that $\traj{}$ remains as close to $\autotraj{}$ as possible.
\newcommand{\subindexset}[2]{\mathcal{I}^{#1}_{#2}}

\section{Method}
\label{sec:method}

We propose a naturalistic projection technique that represents potentially multimodal human behavior data as a \nset{}.
Specifically, we solve the \nset{} generation problem by using a clustering technique to identify the number of clusters $k_t$ at each time and then generating each convex set $\convsubset{t}{i}$ by computing a convex hull over the points in each cluster.
Then, we apply a dynamically consistent mixed-integer optimization algorithm to project arbitrary trajectories into the representation.


\newcommand{\exampledataset}{\mathbb{P}}

\subsection{Preliminaries: The Clustering Problem}
\label{ssec:prelims-clustering}
In our work, clustering serves as a means of identifying discrete modes of behavior among multiple similar but not identical states in trajectories.
Thus, we briefly introduce the clustering problem alongside two specific clustering algorithms that will be used in this work.

Consider a dataset of $\numdatapts{}$ $\numstates{}$-dimensional data points, $\exampledataset{} = \{\state{}^0, \state{}^1, \cdots, \state{}^{\numdatapts{}-1}\} \in \mathbb{R}^{\numstates{}}$.
Clustering techniques partition the data points of $\exampledataset{}$ into $k$ clusters, $\{ \subindexset{0}{\exampledataset{}}, \subindexset{1}{\exampledataset{}}, \cdots, \subindexset{k-1}{\exampledataset{}} \}$.
Each cluster is a subset of indices of the original dataset $\exampledataset{}$, $\subindexset{j}{\exampledataset{}} \subseteq [\numdatapts{}]$, which maximize the similarity of the associated points within the same cluster and minimize the similarity between those in different clusters.
Clustering assigns each data point to one of $k$ clusters, ensuring that they 
are mutually exclusive,
\begin{equation}
    \label{eq:intersection-of-index-sets}
    \emptyset{} = \subindexset{j}{\exampledataset{}} \cap \subindexset{l}{\exampledataset{}}, \qquad \forall j \neq l; j, l \in [k], 
\end{equation}
and that their union contains all indices of the original dataset,
\begin{equation}
    \label{eq:union-of-index-sets}
    [\numdatapts{}] = \subindexset{0}{\exampledataset{}} \cup \subindexset{1}{\exampledataset{}} \cup \cdots \cup \subindexset{k-1}{\exampledataset{}}.
\end{equation}
Clustering
requires us to define a similarity metric between points, $D:\mathbb{R}^{\numstates{}}\times\mathbb{R}^{\numstates{}} \to \mathbb{R}$.
The most common choice for $D$ is the Euclidean distance metric, and our work uses it when clustering unless otherwise, noted.
Other metrics may be used for different purposes, as with the Hausdorff trajectory similarity metric \citep{chen2011clustering}.
We denote the clustering method as
\begin{equation}
    \label{eq:clustering-alg}
    \left(k, \left\{ \subindexset{0}{\exampledataset{}}, \subindexset{1}{\exampledataset{}}, \cdots, \subindexset{k-1}{\exampledataset{}} \right\}\right) = \mathcal{C}(\exampledataset{}; \Theta, \text{D}),
\end{equation}
where $\Theta$ denotes the algorithm parameters, which vary based on which clustering algorithm is chosen.
The exact choice of clustering algorithm and its configuration affect the type of clusters that may be produced, which will have a signficant impact on the results of our method.

\subsubsection{Approaches to Clustering}
\label{sssec:approaches-to-clustering}


The literature includes a wide variety of approaches to clustering, e.g. partition-based methods, hierarchical methods, and density-based methods, among others.
Partitioning methods for clustering typically seek to minimize variance within each cluster.
One category of such methods is $k$-means clustering \citep{lloyd1982kmeans}, which minimizes the sum of the distance between each point and the centroid of its assigned cluser. 
This algorithm iteratively recomputes cluster centroids and then updates the set of clusters until the set of all clusters finds a local minimum.
Thus, it is simple to use and efficient for large datasets.
However, $k$-means clustering requires the number of clusters to be specified in advance, and the parameter $k$ is most often tuned based on domain-specific knowledge or heuristic tuning methods.
While $k$-means guarantees that all points are present in one of the subsequent clusters, it can be sensitive to outliers and result in high concentrations of points in a few clusters and most clusters containing few points.
The latter of these weaknesses are addressed by methods like \cite{levy2018kmeans-means-constrained}, which ensures a minimum number of points per cluster.


Another category of clustering algorithms is hierarchical clustering, which models possible clusters in a tree, where each leaf of the tree is associated with a unique data point and each subtree with a cluster containing all data points of the nodes within it.
Hierarchical clustering algorithms thus begin with all points in one cluster and split them, or they begin with each data point in its own cluster and merge clusters \citep{mllner2011ModernHA}.

A third category of clustering algorithms involve density-based methods, which form clusters from the most densely packed regions based on computing a pairwise distance matrix between points.
These methods can identify arbitrarily shaped clusters, are more robust to noise, and automatically identify the number of clusters $k$.
However, they tend to suffer from sensitivity to the choice of parameters, have higher computational complexity due to the need to compute the distance matrix, and struggle with high-dimensional data.
\texttt{HDBSCAN} (Hierarchical Density-Based Spatial Clustering of Applications with Noise \cite{campello2015hierarchical}), which uses techniques from both hierarchical and density-based clustering, computes a spanning tree and then breaks it into clusters using the distance matrix.
It can handle clusters of varying densities more robustly than other density-based clustering algorithms.
\hmzh{\texttt{HDBSCAN} identifies outliers by placing them into a special ``extra'' cluster.
Ignoring this cluster in downstream tasks thus becomes an outlier rejection method.
Thus, \texttt{HDBSCAN} does not guarantee, as does $k$-means clustering, that every data point ends up in a valid cluster---those that \texttt{HDBSCAN} deems noisy may be dropped.}
\texttt{HDBSCAN} also allows the end user to specify a minimum number points per cluster, and
although the distance metric used in these algorithms can be configured depending on the application, the Euclidean distance metric is often used.

\newcommand\trajhorizon[1]{H^{#1}}
\newcommand\newdataset{\tilde{\dataset{}}}
\newcommand\hullsubstatefn[2]{\hullstate{}(\state{#2}^{#1})}

\newcommand{\startset}{\mathcal{S}}
\newcommand{\finalset}{\mathcal{E}}

\subsection{\NSET{} Identification}
\label{ssec:method-naturalistic-set}

\subsubsection{Dataset Generation}
\label{sssec:method-dataset}
Based on \Cref{ass:historical-data}, we assume access to a dataset as defined in \eqref{eq:dataset}, $\dataset{} = \{ \traj{i} \}_{i\in [\numdatapts{}]}$, consisting of $\numdatapts{}$ trajectories of length $\horizon$.
From $\dataset{}$, we construct a collection of time-indexed subdatasets $\{ \subdataset{t} \}$, one per time step $t \in [\horizon{}+1]$, by aggregating all states at time $t$ across the trajectories in $\dataset{}$,
\begin{equation} 
\label{eq:dataset-slice-original}
\subdataset{t} = \left\{ \state{t}^{0}, \state{t}^{1}, \ldots, \state{t}^{\numdatapts{}-1} \right\}.
\end{equation}
In some applications, an end user may prefer to use a transformed, perhaps lower-dimensional, version of the state.
To this end, we introduce $y(\bm{x})$, an optional transformation of the state $\bm{x}$ that enables a user to transform the state over which we compute the convex sets.
Applying this transformation to the subdataset yields
\begin{equation} 
\label{eq:dataset-slice}
y\left(\subdataset{t}\right) = \left\{ y\left(\state{t}^{0}\right), y\left(\state{t}^{1}\right), \ldots, y\left(\state{t}^{\numdatapts{}-1}\right) \right\},
\end{equation}
for all $t \in [\horizon{}+1]$, from which we seek to learn $\wptset{t} \in \natset{}{}$.
We call these transformed states, which have dimension $\numhullstates{} \leq \numstates{}$, \emph{hull states} (for reasons that will become clear) and denote them $\hullstate{t}^i = y(\state{t}^i) \in \mathbb{R}^{\numhullstates{}}$.

Recall that a nonconvex set can be represented as a union of convex sets, per \cref{eq:nat-subset-at-t}.
Thus, at each time $t$, our goal is to use the subdataset \cref{eq:dataset-slice} to estimate the corresponding nonconvex subset $\wptset{t} \in \natset{}{}$ by identifying the number of convex sets needed to represent the behavior at time $t$, $\numconvexsets{t}$, and the convex sets themselves, $\{ \convsubset{t}{0}, \convsubset{t}{1}, \ldots, \convsubset{t}{\numconvexsets{t}-1} \}$.

\subsubsection{Identifying Time-Indexed Discrete Behavior Modes using Clustering}
\label{sssec:clusters-of-convex-sets}
Next, we present our approach to solving the first subproblem of naturalistic set generation.
Specifically, to construct the time-indexed subset at each time $t$, $\wptset{t}$, we must identify the number of convex sets within $\wptset{t}$, $k_t$.
Our method identifies $k_t$ and also the mutually exclusive index sets $\{ \subindexset{i}{y(\subdataset{t})} \}_{i\in [k_t]}$ with a clustering approach as discussed in \cref{ssec:prelims-clustering}.
Let $\mathcal{C}_t$ be a clustering algorithm chosen by the user and $\Theta_t$ its associated parameters, both at time $t$.
Then, we apply \cref{eq:clustering-alg} to compute the clusters, so
\begin{equation}
    \label{eq:clustering-in-method}
    \left( k_t, \{ \subindexset{i}{y(\subdataset{t})} \}_{i\in [k_t]} \right) = \mathcal{C}_t(y(\subdataset{t}); \Theta_t, \text{d}).
\end{equation}
Using different clustering algorithms results in different types of representations of naturalistic behavior, as we discuss in more detail in our experiments in \cref{sec:experiments}.
We rely on the user to specify parameters $\Theta_t$ for each clustering algorithm appropriately based on the application---configuration details like these can affect how many clusters are made and how many points are in each cluster, among other effects.

\subsubsection{Computing Convex Sets}
\label{sssec:computing-convex-sets}
Once we identify the number of convex sets, $k_t$, to be generated from $\subdataset{t}$, we must address the second subproblem of naturalistic set generation, generating $k_t$ convex sets from subdataset $y(\subdataset{t})$.
We do so by computing convex hulls over each cluster.
Recall that, in addition to identifying $k_t$, clustering associates each data point in $y(\subdataset{t})$ with a  cluster $\subindexset{i}{y(\subdataset{t})}$.
Thus, we construct each convex set $\convsubset{i}{t}$ within $\wptset{t}$ by computing the convex hull of the data points associated with cluster $\subindexset{i}{y(\subdataset{t})}$, specifically
\begin{equation}
    \label{eq:convex-hull}
    \convsubset{t}{i} 
    = \text{ConvexHull}\left(\left\{ \hullstate{t}^{j} ~|~ j \in \subindexset{i}{y(\subdataset{t})} \right\}\right).
\end{equation}
The naturalistic behavior subset $\wptset{t}$ can then be constructed from these convex hulls, as in \cref{eq:nat-subset-at-t}.
Finally, the \nset{} $\natset{}{}$ can be constructed as a collection of subsets containing multiple convex hulls indexed in time, as in \cref{eq:full-nat-set}.

By representing \nsets{} as collection of convex hulls, we benefit from several desirable qualities.
\begin{enumerate}
    \item \textbf{Interpretability}: A given facet of the convex hull exists because it connects two or more points of observed naturalistic data.
    As such, convexity implies that an arbitrary point in the hull sits within boundaries established by the naturalistic data.

    \item \textbf{Minimal Assumptions in Modeling Human Behavior}: In contrast with generative modeling approaches like \cite{bhattacharyya2023generative} that have been proposed for modeling human behavior, our approach does not assume the existence of any particular probability density over the set of naturalistic trajectories.
    This removes a potential source of imprecision.

    \item \textbf{Data Efficiency}: Using convex hulls avoids requiring a large amount of data.
    Thus, our method provides the benefit of working on smaller datasets, as compared to more data-intensive learning methods.

    \item \textbf{Computability}: Convex hulls can be computed efficiently.
    In general, the worst-case computational complexity of producing a convex hull from $\numdatapts{}$ points in $\mathbb{R}^\numhullstates{}$ is $O(\numdatapts{}^{\lfloor \numhullstates{}/2\rfloor})$ \citep{barber1996quickhull}.
    For lower dimensions ($\numhullstates{} \leq 3$), a convex hull can be computed in $O(\numdatapts{}\log\numdatapts{})$ \citep{barber1996quickhull}. 

    \item \textbf{Transferrability of Representation}: A convex hull can be equivalently represented by affine inequality constraints,
    \begin{equation}
    \label{eq:half-space-intersection}
    G^i_t \hullstate{t} \leq h^i_t,
    \end{equation}
    where $(G^i_t, h^i_t) \in \mathbb{R}^{n_f\times\numhullstates{}} \! \times \mathbb{R}^{n_f}$ defines a convex polytope formed by the $n_f$ facets of the $\numhullstates{}$-dimensional hull.
    Expressing the naturalistic behavior set in this form allows us to utilize \cref{eq:half-space-intersection} as a constraint within an optimization problem.
\end{enumerate}

\subsubsection{Additional Considerations}
\label{sssec:additional-considerations}

In this section, we make note of a few technical considerations in generating \nset{}s with our method.
First, computing a convex hull requires a minimum of $d+1$ points in a $d$-dimensional space.
Thus, the clustering algorithm must be chosen carefully to ensure that this constraint is satisfied.

Next, the number of facets, $n_f$, on the boundary of a convex hull can rise exponentially with $\numhullstates{}$.
As each facet corresponds to a half-space, introducing additional facets increases the number of constraints required to represent a convex polytope as a half-space intersection (i.e., affine inequality constraints).
To address this problem, we can use the transformation $y(\cdot)$ introduced in \cref{eq:dataset-slice} to reduce the dimensionality of the problem and limit the impact of this issue.
In general, having few hull states (i.e., small $\numhullstates{}$) results in a computationally efficient algorithm.
However, if more hull states are needed, polynomial-time approximation algorithms for computing convex hulls do exist \citep{Sartipizadeh2016ComputingTA,balestriero2022deephull}.
We further discuss these considerations regarding our choice to build convex hulls using clustered data points in \cref{sec:limitations}.

\subsubsection{Computational Runtime (Offline)}
\label{sssec:computational-runtime}
To generate a \nset{} $\natset{}{}$, our method requires generating $\horizon{}$ naturalistic subsets, one at each time.
Generating the subset $\wptset{t}$ requires running a clustering algorithm and a convex hull computation in sequence.
For the sake of analysis, we consider Lloyd's algorithm for $k$-means clustering, which has worst-case time complexity $O(\numdatapts{} ~ k_t ~ \numhullstates{} )$ assuming a constant, predetermined number of iterations \citep{lloyd1982kmeans}.
Once we have identified the clusters, we generate a convex hull for each of the $k_t$ clusters, which requires $O(k_t ~ \numdatapts{}^{\lfloor \numhullstates{}/2 \rfloor})$.
We expect that $k_t \ll \numdatapts{}$, so let $K = \max_t k_t$ be the largest number of clusters, which can be used to define the time complexity.
For $\numhullstates{} \geq 4$, convex hull generation will take longer, and so, the overall computational complexity of \nset{} generation is $O(\horizon{} ~ K ~ \numdatapts{}^{\lfloor \numhullstates{}/2\rfloor})$.
For a different clustering algorithm with complexity $H(\numdatapts{}, k_t, \numhullstates{})$, the runtime would be $O(\horizon{} \cdot \max{H(\numdatapts{}, K, \numhullstates{}), k_t ~ \numdatapts{}^{\lfloor \numhullstates{}/2 \rfloor}})$.
\newcommand{\optflags}[2]{s^{#1}_{#2}}
\newcommand{\optflagvec}[1]{\mathbf{s}_{#1}}
\newcommand\largeno{S}

\subsection{Projection Into the \NSET{}}
\label{ssec:method-projection}

In this section, we describe a mixed-integer optimization problem that, when solved, allows us to project an autonomously planned trajectory, $\autotraj{}$, into \nset{} $\natset{}{}$, making it more naturalistic while maintaining dynamic feasibility.
This projection problem involves two subproblems. First, we seek to identify a single tube within the naturalistic set, as described by the right-hand side of \cref{eq:naturalistic-trajectory-implication}.
Second, we seek to project $\autotraj{}$ into this tube.
We note that \cite{khan2024actnatural} solves the second subproblem using an optimization-based projection.
In this work, we simultaneously solve both subproblems by extending the optimization defined in Eq. (9) of \cite{khan2024actnatural} with binary variables. 

Given a \nset{} $\natset{}{}$ generated as described in \cref{sssec:clusters-of-convex-sets}, we seek to identify a set of controls $\ctrl{0}, \ldots, \ctrl{\horizon-1}$ that generates a naturalistic trajectory $\traj{}$ similar to $\autotraj{}$  and subject to the dynamic constraints $\dynamics{}$ in \cref{eq:dynamics}.
We also seek to identify binary variables at each time $t$,
\begin{equation}
    \label{eq:binary-flags-def}
    \optflagvec{t} = \left[\optflags{0}{t} ~ \optflags{1}{t} ~ \cdots ~ \optflags{k_t-1}{t}\right]^\intercal \in \{0, 1\}^{k_t},
\end{equation}
which indicate which convex set $\convsubset{t}{j}$ we will be projecting $\autotraj{}$ into at time $t$.
Let $\state{init}$ be the initial condition at time $t=0$ of $\xi_{a}$.
We define the projection as a mixed-integer problem,
\begin{subequations}
\label{eq:projection-opt}
\begin{align}
\min_{\substack{\ctrl{0}, \ldots, \ctrl{\horizon-1},\\ \optflagvec{0}, \optflagvec{1}, \ldots, \optflagvec{\horizon{}}}} \quad & \text{D}(\autotraj{}, \traj{}) + \gamma \sum_{t=0}^{\horizon{}-1} \text{L}(\ctrl{t}) \label{eq:projection-opt-objective} \\
    \textrm{s.t.} \quad & \traj{} = [ \state{0}^\intercal ~ \state{1}^\intercal ~ \cdots ~ \state{\horizon}^\intercal ]^\intercal \label{eq:projection-opt-traj} \\
              & \state{t+1} = \dynamics{}(\state{t},  \ctrl{t}) \quad \forall t \in [\horizon{}+1] \label{eq:projection-opt-dynamics} \\
              & \state{0} = \state{init} \\
              & y(\state{t}) \in \wptset{t} ~\!\quad\quad\quad \forall t \in [\horizon{}+1],
              \label{eq:projection-opt-nset}
\end{align}
\end{subequations}
where $D : \trajspace{} \times \trajspace{} \to \mathbb{R}$ in \eqref{eq:projection-opt-objective} is a distance metric for trajectories, e.g. the Euclidean distance or a trajectory similarity metric \citep{chen2011clustering} and L$(\cdot)$ defines a regularization on the control variables.
The constraints in \eqref{eq:projection-opt-dynamics} and \eqref{eq:projection-opt-nset} enforce dynamic feasibility and the learned naturalistic behavior constraints.
Using standard techniques in mathematical programming, we can rewrite \cref{{eq:projection-opt-nset}} using two constraints.
Let $\largeno{} \in \mathbb{R}$ be a large number.
Then, \cref{eq:projection-opt-nset} at each timestep $t$ can be rewritten as 
\begin{subequations}
\label{eq:natset-lin-inequality}
\begin{align}
    G^j_t y_t &\leq h^j_t + (1 - \optflags{j}{t}) \largeno{} \mathbb{1}^{n_f}  \quad \forall j \in \left[k_t\right], \label{eq:linear-or-inequality} \\
    \!\sum_{j=0}^{k_t-1} \optflags{j}{t} &\geq 1 \quad\quad\quad\quad\quad\quad\quad\quad~ \forall j \in \left[k_t\right]. \label{eq:binary-sum}
\end{align}
\end{subequations}

Eq. \cref{eq:linear-or-inequality} utilizes the equivalent representation of an individual convex hull $\convsubset{t}{j}$ as a set of linear half-space inequalities as decribed by \eqref{eq:half-space-intersection}.
To avoid the nonsensical requirement that the hull state $y_t$ be in every convex hull at each time, we modify the inequality to only enforce the constraint when $\optflags{j}{t}$ is enabled. 
To ensure that the hull state is within at least one convex hull, we require in the second constraint that the value of the binary flags sums to at least one.
In addition, we note that we can easily augment \eqref{eq:projection-opt} to enforce additional constraints such as control limits or safety restrictions.

\begin{remark}[Modifying \cref{eq:binary-sum} to Improve Computational Efficiency]
\label{rmk:tighten-feasible-region}
The definition of naturalistic trajectories in \cref{eq:naturalistic-constraint} suggests that a state may be in multiple convex hulls at a given time $t$. 
In practice, many clustering algorithms produce disjoint clusters. Convex hulls computed over such clusters may not overlap, and in these cases it can be more computationally efficient to change the inequality in \cref{eq:binary-sum} to an equality constraint.
\end{remark}

The formulation in \cref{eq:projection-opt} is a mixed-integer problem, unlike that of \cite{khan2024actnatural}.
As with that work, the projection in \eqref{eq:projection-opt} may also be non-convex if the dynamics $\dynamics{}$ are nonlinear.
Nevertheless, a variety of well-studied techniques exist to identify local minimizers of mixed-integer non-convex problems like \eqref{eq:projection-opt}, using techniques like branching and bounding \citep{land2010branchbound}.
We refer the reader to \cite{nocedal1999numerical, floudas1995nonlinear} for further details about solving continuous and mixed-integer optimization problems. 

\newcommand\movingvehspace{\mathcal{V}}
\newcommand\mass{M}

\section{Experiments}
\label{sec:experiments}

\topic{We demonstrate our multimodal naturalistic projection technique on real-world human driving data.} 

\subsection{The {inD} and {rounD} Datasets}
\label{ssec:inD-dataset}
For this work, we utilize the {inD} and {rounD} naturalistic datasets \citep{inD2020naturalisticdataset, rounDdataset}, which record and label driving data for cars, bicyclists, and pedestrians at German intersections and roundabouts from a camera positioned above the scene.

Each actor $i$'s trajectory is annotated at time $t$ with state \
\begin{equation}
\label{eq:inD-state}
\stategt{t}{i} = [p_t^{i\intercal} ~ v_t^{i\intercal} ~ a_t^{i\intercal} ~ \theta_t ]^\intercal,
\end{equation}
containing planar position $p_t^i$ tracking the center of the actor, planar velocity $v_t^i$, planar acceleration $a_t^i$, and heading $\theta_t^i$.
Trajectory $\xi^i$ for actor $i$ is constructed as in \eqref{eq:trajectory-form} and states are sampled at $25$ frames per second, with actor $i$ being visible and recorded from the first frame in which actor $i$ is visible, at $t=0$, 
until the last frame in which it is visible, at $t=\trajhorizon{i}$.

\subsection{Curating the Naturalistic Dataset}
\label{ssec:experiments-dataset}

We generate a multimodal dataset $\dataset{}$ from the inD and rounD data by heuristically filtering out scenarios in which moving vehicles (defined by index set $\movingvehspace{}$) begin within starting set $\startset{}$ at $t = 0$ and end in final set $\finalset{}$ at time $t = \horizon{}$. 
This filter is defined by the following indicator function:
\begin{equation}
\label{eq:exp-heuristic}
h(\xi^i; \startset{}, \finalset{}) = i \in \movingvehspace{} ~\wedge~ \state{0}^i \in \startset{} ~\wedge~ \state{\horizon}^i \in \finalset{}.
\end{equation}
To ensure nontrivial behavior, the first term of \eqref{eq:exp-heuristic} considers only moving vehicles.
The second term further filters the naturalistic trajectories under the assumption that every actor moving from a start set $\startset{}$ (shown in \cref{fig:ind-r1} and \cref{fig:exp-3-round-r0} in green to some end set $\finalset{}$ (shown in \cref{fig:ind-r1}).
By setting $\finalset{}$ to contain an entire intersection (and thus become trivial), we demonstrate an example where only $\startset{}$ is used to filter tasks.
In such an example, the trajectories may diverge significantly as they progress, and our method handles this divergence appropriately. 
We specify different sets $\startset{}, \finalset{}$ for each experiment.

\subsection{Generating the \NSET{}}
\label{ssec:exp-nset-generation}


Our method requires naturalistic data over which we can compute convex hulls, so we define the hull state using the information available in \eqref{eq:inD-state}.
We model $\dynamics{}$ with planar double-integrator dynamics, where each moving vehicle actor is a point with mass $\mass$ evolving according to 
\begin{equation}
\label{eq:exp-planar-di-dynamics}
\state{t+1} 
= \left[ \begin{array}{c}
    p_{x,t+1} \\
    v_{x,t+1} \\
    p_{y,t+1} \\
    v_{y,t+1} \\
\end{array} \right]
= \left[ \begin{array}{c}
    p_{x,t} + \Delta t v_{x,t} \\
    v_{x,t} + \Delta t F_{x,t} / \mass \\
    p_{y,t} + \Delta t v_{y,t} \\
    v_{y,t} + \Delta t F_{y,t} / \mass \\
\end{array} \right],
\end{equation}
where $\ctrl{t} = [F_{x, t} ~ F_{y, t}]^\intercal$ is the force applied to the point mass at time $t$.
As \eqref{eq:exp-planar-di-dynamics} constitutes a linear equation in $\state{t}$ and $\ctrl{t}$, we denote the dynamics as $\state{t+1} = A\state{t} + B\ctrl{t}$ for brevity, where $A \in \mathbb{R}^{\numstates{} \times \numstates{}}, B \in \mathbb{R}^{\numstates{}\times \numctrls{}}$\!\!.
In practice, many systems of interest---including vehicle models---are differentially flat and admit a representation of state and control in which dynamics are linear \citep[Ch. 9]{sastry2013nonlinear}.

Next, we define the hull state by extracting the two-dimensional position from state $\state{t}$ of the dynamics in \eqref{eq:exp-planar-di-dynamics},
\begin{equation}
\label{eq:exp-hullstate}
\hullstate{t} = \hullsubstatefn{}{t} = \left[ \begin{array}{cccc}
    1 & 0 & 0 & 0  \\
    0 & 0 & 1 & 0 
\end{array} \right]  \state{t}.
\end{equation}
As the full \nset{} $\natset{}{}$ captures sets of positions over time, we neglect higher order kinematics and assume these states are observable from position, although they can be included in principle at the expense of additional computation.
For example, \eqref{eq:inD-state} includes velocity and acceleration, so we could use planar triple-integrator dynamics with jerk-based controls for experiments on this dataset.

Using \eqref{eq:exp-hullstate}, we generate subdatasets $\{ \subdataset{t} \}_{t=0}^{\horizon{}}$ as described by \eqref{eq:dataset-slice}.
At each timestep, we run a clustering algorithm to identify $k_t$ clusters of points, $\{ \mathcal{C}^{j}_{t} \}_{j=0}^{k_t-1}$.
We use two clustering algorithms in our experiments, \kmeans{} and \texttt{HDBSCAN}, so that we can compare the differences in results between the two algorithms with the baseline unimodal approach from \cite{khan2024actnatural}.
Upon clustering subdataset $\subdataset{t}$, we generate the naturalistic subset, $\wptset{t}$, as per \cref{eq:nat-subset-at-t} by computing the convex hull on each cluster with the Quickhull algorithm \citep{barber1996quickhull} as described by \eqref{eq:convex-hull}.

As a convex hull can only be generated from a subdataset with at least $\numhullstates{}+1$ points,
we introduce mechanisms to ensure that each cluster $\mathcal{C}^i_t$ at time $t$ has enough points to form a convex hull, $|\mathcal{C}^i_t| \geq \numhullstates{}+1 = 3$.
Unimodal naturalistic set generation.
For $k$-means clustering, we enforce this requirement by using $k$-means-constrained clustering \cite{levy2018kmeans-means-constrained}, which modifies $k$-means using a min-cut algorithm to ensure each cluster contains sufficient points at the expense of an increase in computational complexity.
Meanwhile, \texttt{HDBSCAN} allows the direct configuration of this quantity, and so we set it to 3 for our experiments.
Moreover, we select the maximum distance parameter to $\epsilon=1$ meter, ensuring that points closer to one another than $\epsilon$ are included in the same cluster.
\hmzh{Lastly, we neglect to include the ``noisy'' cluster (described in \cref{ssec:prelims-clustering}) in the \nset{}.}

Finally, we build the full \nset{} $\natset{}{}$ as in \eqref{eq:full-nat-set}, using horizon $\horizon{}$ which is the maximum time for which all clusters satisfy the size condition.

\newcommand\autohorizon{\trajhorizon{}_a}
\subsection{Framing the Projection Problem}
\label{ssec:exp-projection}
For the subsequent experiments, we frame the projection of $\autotraj{}$ into $\natset{}{}$ as an alteration of \eqref{eq:projection-opt},

\begin{subequations}
\label{eq:projection-opt-exp}
\begin{align}
\min_{\substack{\ctrl{0}, \ldots, \ctrl{\autohorizon{}\!\!-1},\\ \optflagvec{0}, \optflagvec{1}, \ldots, \optflagvec{\autohorizon{}}}} \quad & \| \autotraj{} - \traj{} \|_2^2 + \gamma\| \ctrl{0:\autohorizon{}\!\!-1} \|_2^2  \label{eq:projection-opt-objective-exp} \\
    \textrm{s.t.} \quad & \traj{} = [ \state{0}^\intercal ~ \state{1}^\intercal ~ \cdots ~ \state{\autohorizon}^\intercal ]^\intercal \label{eq:projection-opt-traj-exp} \\
              & \state{t+1} = A\state{t}+ B\ctrl{t} \quad\quad\quad\quad \forall t \in [\autohorizon{}] \label{eq:projection-opt-dynamics-exp} \\
              & \state{0} = \state{init} \label{eq:init-cond-exp} \\
              & \!G^j_t y_t \leq h^j_t + (1 \!-\! \optflags{j}{t}) \largeno{} \mathbb{1}^{n_f} \!\!\! \left. \begin{array}{c}
                \forall j \in [k_t] \\
                \forall t \in [\horizon{}\!+\!1]
              \end{array} \right.\!\!\!
              \label{eq:projection-opt-nset-exp} \\
              & \!\sum_{j=0}^{k_t-1} \optflags{j}{t} = 1, \!\quad\quad\quad\quad\quad\quad \forall t \in [\horizon{}\!+\!1] , \label{eq:at-least-1-exp}
\end{align}
\end{subequations}
where $\autotraj{}$ has horizon $\autohorizon{} \in \mathbb{N}$ and $\horizon = |\natset{}{}|$.
Equation \eqref{eq:projection-opt-objective-exp} defines the similarity objective $D(\cdot,\cdot)$ \hmzh{and the control cost $L(\cdot)$ as the Euclidean distance, with $\gamma=0.1$.}
Equation \eqref{eq:projection-opt-dynamics-exp} enforces (linear) planar double-integrator dynamics over the entire trajectory horizon $\autohorizon{}$ as described by \eqref{eq:exp-planar-di-dynamics}. 
Note that we adjust the projection to account for the case where $\horizon \neq \autohorizon{}$ which can arise, e.g., when a planned trajectory extends beyond times for which we have behavior data.
Eq. \cref{eq:init-cond-exp} requires that the initial condition is satisfied.
Eqs. \eqref{eq:projection-opt-nset-exp} and \cref{eq:at-least-1-exp} describe the multiple possible naturalistic behavior constraints as linear inequalities, as described by \eqref{eq:natset-lin-inequality}, and we use the modification described in \cref{rmk:tighten-feasible-region} to improve computational efficient by reducing the feasible space.

We note that the objective \cref{eq:projection-opt-objective-exp} is convex and constraints \cref{eq:projection-opt-dynamics-exp} and \cref{eq:projection-opt-nset-exp} are linear, indicating that \eqref{eq:projection-opt-exp} is a (generally NP-Hard) mixed-integer convex optimization problem. 
For this reason, we use the convex optimization library, CVXPY \citep{diamond2016cvxpy} with the Gurobi optimizer \citep{gurobi} to quickly solve \cref{eq:projection-opt-exp}.
The projection problem is feasible as long as $\mathcal{N}$ includes a dynamically feasible trajectory.
Thus, a solution could be guaranteed if, for example, at least one of the trajectories used to form the \nset{} is dynamically feasible.
\hmzh{Moreover, constraints like actuator limits can be added to \cref{eq:projection-opt-exp} if desired.}
In \cref{ssec:exp-timing}, we report the method's runtime 
and a way to trade off projection quality with runtime.

\subsection{Experiment 1: Curved Road}
\label{ssec:exp1-curved-road}

    

\begin{figure*}
\centering
\begin{multicols}{2}
    \begin{subfigure}[t]{0.4\textwidth}
        \centering
        \includegraphics[width=0.7\textwidth]{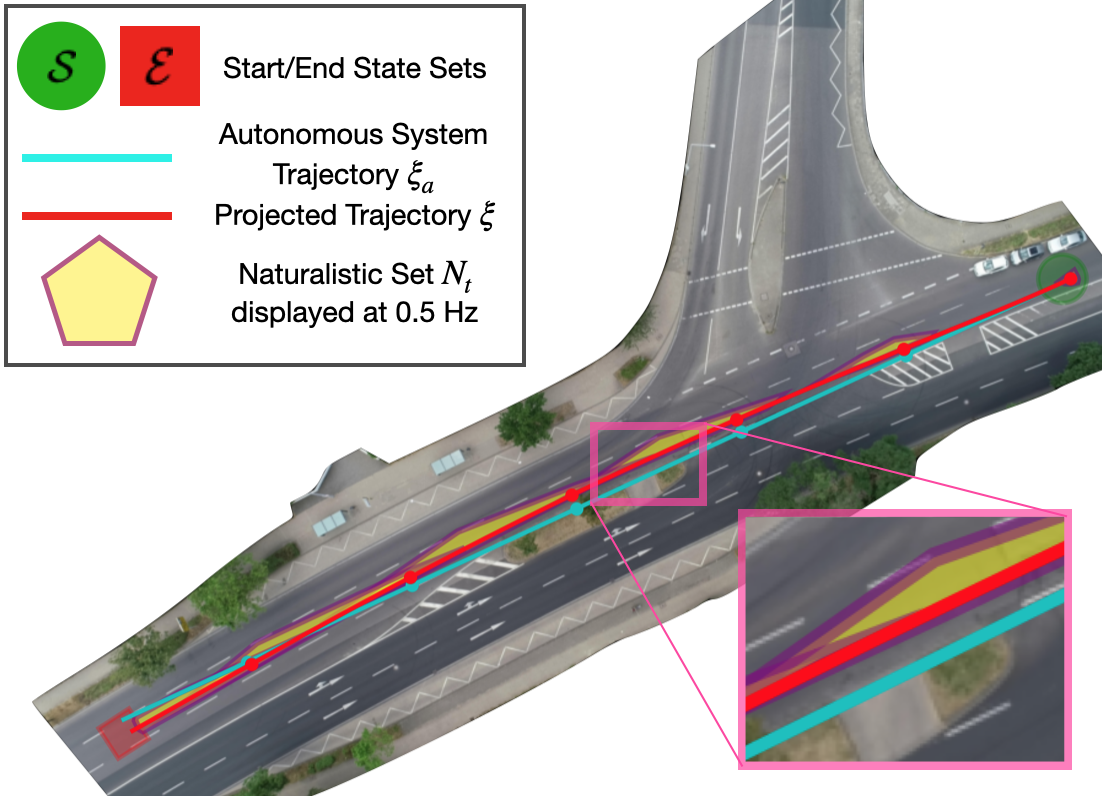}
        \caption{Unimodal naturalistic set generation and projection \cite{khan2024actnatural}.}
        \label{fig:inD-r1-hull2d}
    \end{subfigure}
    \begin{subfigure}[t]{0.4\textwidth}
        \centering
        \includegraphics[width=0.7\textwidth]{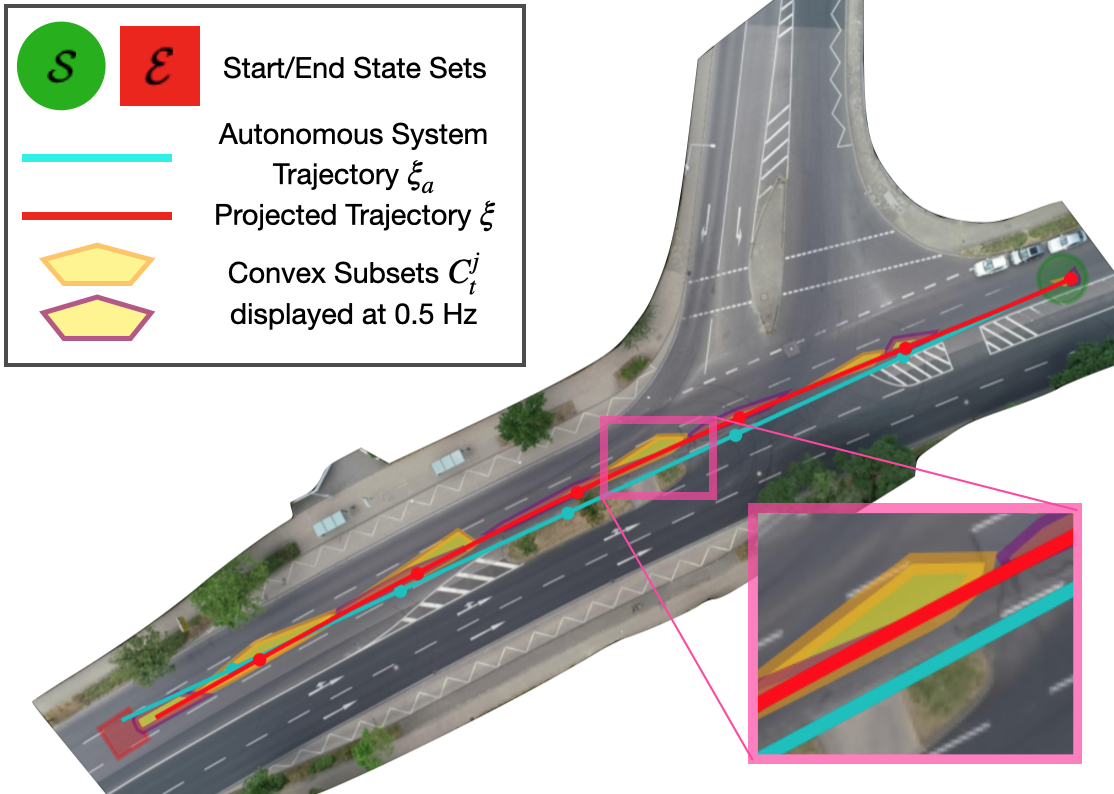}
        \caption{Multimodal naturalistic set generation and projection using 2-means-constrained clustering.}
        \label{fig:inD-r1-clusters2d}
    \end{subfigure}
\end{multicols}
\vspace{-12pt}
\caption{\eqref{fig:inD-r1-hull2d} and \eqref{fig:inD-r1-clusters2d} depict \nset{}s generated using two different methods for recording 1 of the inD dataset. Both methods capture the tendency of human drivers to drive on the outside of the curve \emph{without explicit modeling}.
Projecting a straight trajectory into the \nset{} results in similar trajectories with both methods, that capture these human tendencies.
\label{fig:ind-r1}}
\end{figure*}
\begin{figure*}
\centering
\begin{multicols}{3}
    \begin{subfigure}[t]{0.95\columnwidth}
        \centering
        \vspace{-2em}
        \includegraphics[width=\columnwidth,height=10.85em]{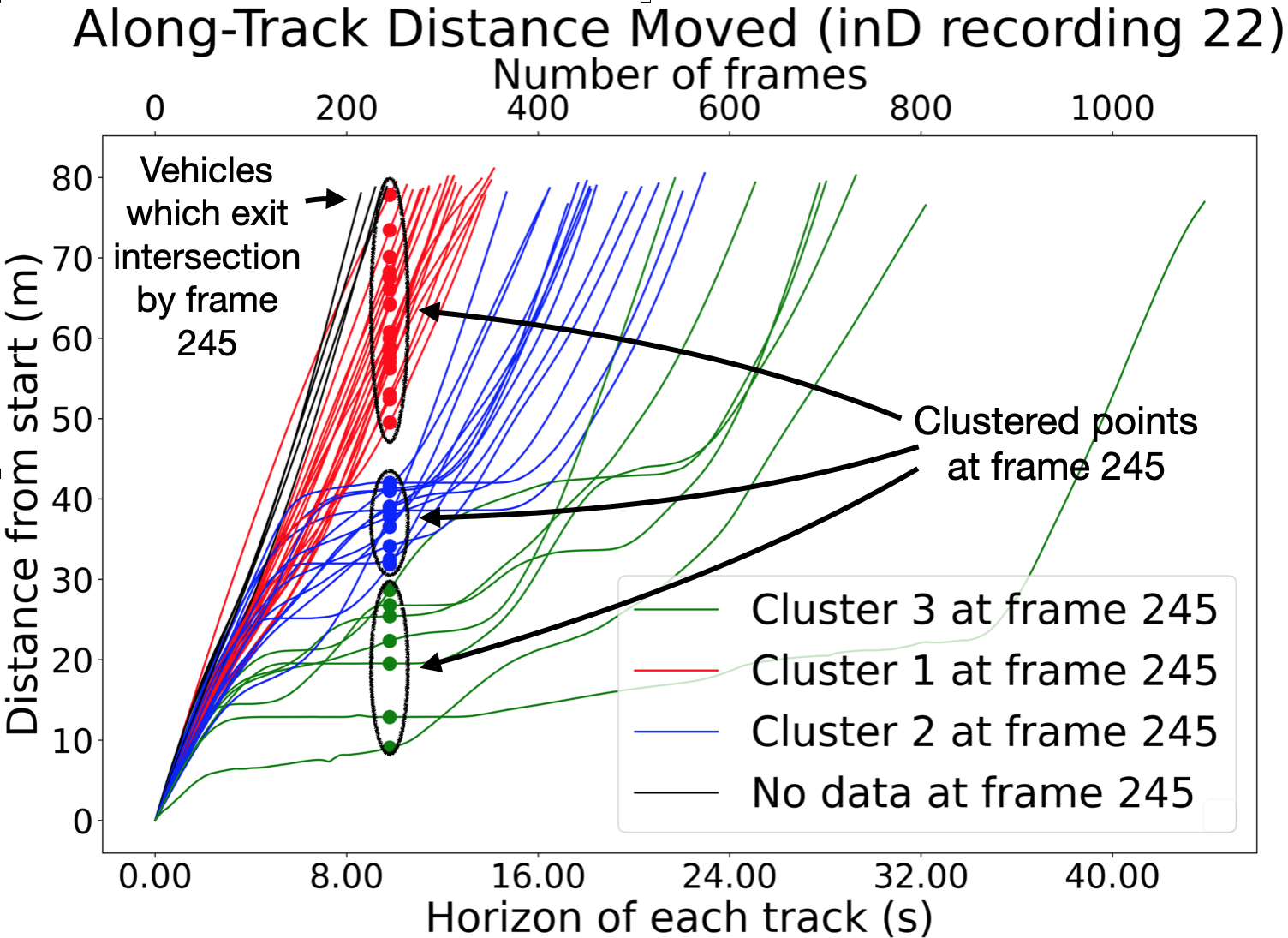}
        \subcaption{Along-track distance plotted against time for the filtered data.}
        \label{fig:inD-r22-trajectory-plots}
    \end{subfigure}

    \begin{subfigure}[t]{\columnwidth}
        \centering
        \includegraphics[width=\columnwidth,height=10em]{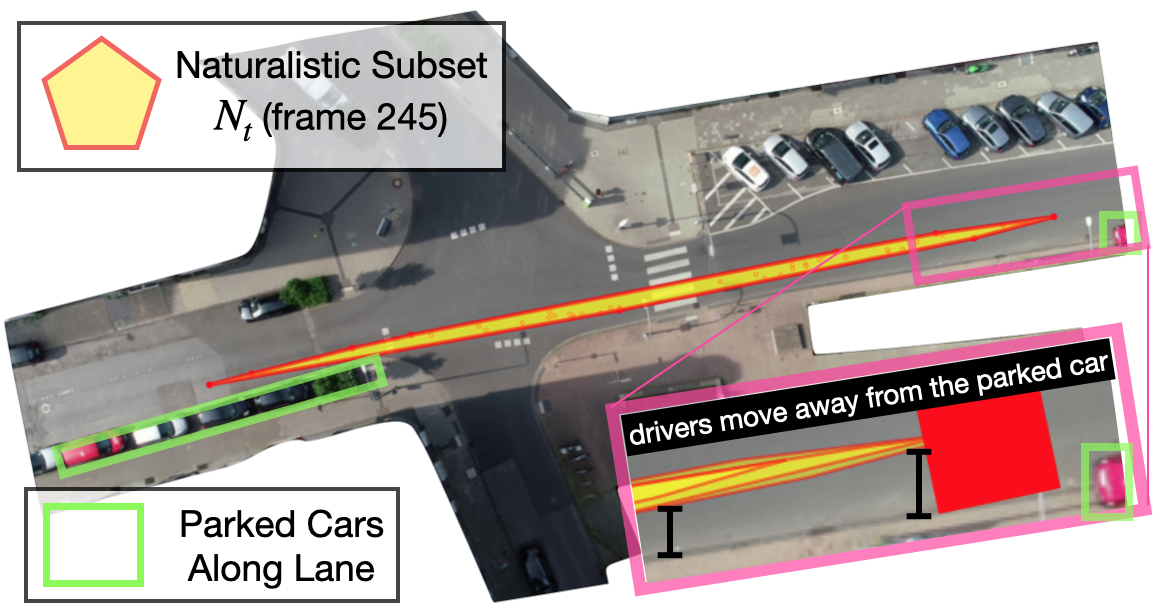}
        \subcaption{\Nset{}s generated at frame 245 using \citep{khan2024actnatural}.
        }
        \label{fig:inD-r22-hull2d-f245}
    \end{subfigure}
    
    \begin{subfigure}[t]{\columnwidth}
        \centering
        \includegraphics[width=\columnwidth,height=10em]{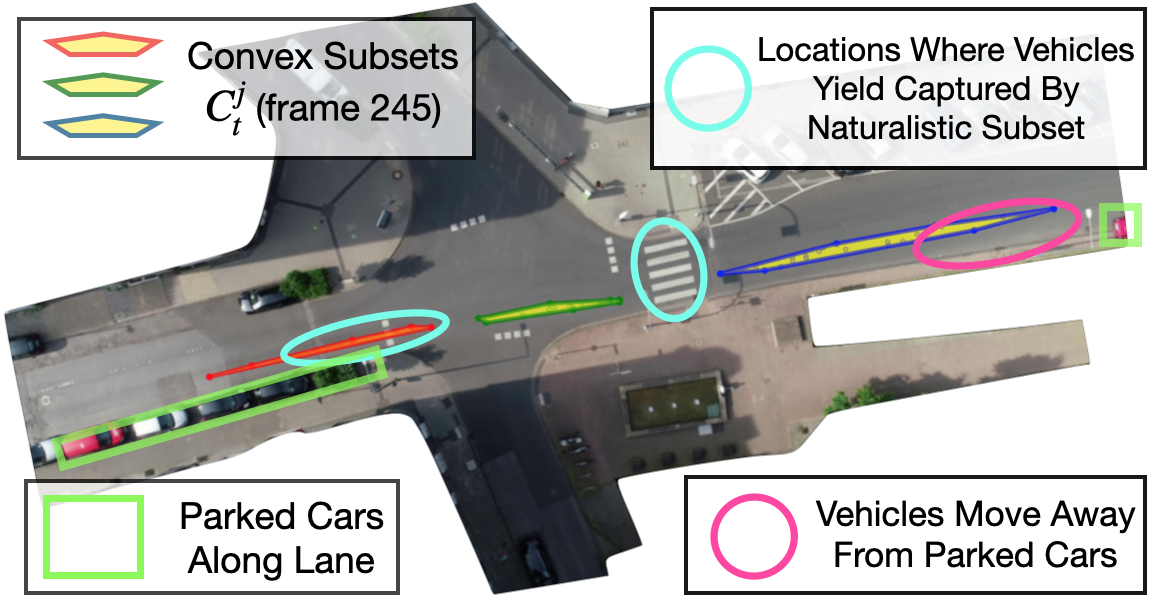}
        \subcaption{\Nset{}s generated at frame 245 using multimodal naturalistic set generation.}
        \label{fig:inD-r22-clusters2d-f245}
    \end{subfigure}
\end{multicols}
\vspace{-12pt}
\caption{An analysis of the behavior of vehicles traveling eastward along the road in recording 22 of the inD dataset, without explicit modeling.
\eqref{fig:inD-r22-hull2d-f245} depicts a large set spanning most of the relevant path and which may not be as useful for understanding human-like behavior.
The method used in \eqref{fig:inD-r22-clusters2d-f245} breaks up the \nset{} and exposes areas where vehicles tend to slow or stop for other agents.
We identify the number of clusters by inspecting \cref{fig:inD-r22-trajectory-plots}.
\label{fig:inD-r22}}
\vspace{-1.5em}
\end{figure*}

\cref{fig:ind-r1} depicts two curved roads separated by a median running through a T-intersection.
We define $\dataset{}$ to include the positions of all moving vehicles whose trajectories begin and end in the second lane of the upper road, where $\startset{}$ is indicated by the green circle and $\finalset{}$ by the red square. 
Filtering based on these criteria, based on \cref{eq:exp-heuristic}, results in 32 trajectories.
We will show that the \nset{} reflects how vehicles are influenced by the curve of the road, without explicitly modeling their preferences.

\subsubsection{Configuration}
In this experiment, we use 2-means-constrained clustering. The resulting \nset{} is generated over the horizon $\horizon{} = 220$.

\subsubsection{Analysis}
The unimodal and multimodal \nset{}s generated from $\dataset{}$ are shown at frames 0, 40, 80, 120, 160, and 200 in \cref{fig:inD-r1-clusters2d,fig:inD-r1-hull2d}.
In \cref{fig:inD-r1-hull2d}, the naturalistic sets $\{\wptset{t}\}$ begin compact but lengthen along the lane over time, indicating that vehicles drive at different speeds along this lane.
In \cref{fig:inD-r1-clusters2d}, we note that the vehicles that drive more slowly are clustered into the rear hull (which contains a large majority of the data).
These vehicles drive along a thinner portion of the road, especially from frames 40 to 160, distinguishing them from the front cluster which contains a few trajectories that speed ahead or drive on the sides of the lane.
Specifically, this divergence occurs near the T-intersection, which indicates that many cars drive cautiously when approaching it despite having the right of way.
Additionally, in both figures the \nset{} covers the outside of the most curved portion of the lane but not the inside.
This suggests that drivers naturally hug the outside of a curved lane.

\subsubsection{Projection}
\cref{fig:inD-r1-hull2d,fig:inD-r1-clusters2d} depict $\autotraj{}$, a constant-velocity trajectory moving straight through the second lane.
This trajectory is not naturalistic, i.e. it is not within the \nset{} according to \cref{eq:naturalistic-constraint}.
In particular, we note that it gets abnormally close to the median, which could negatively impact safety and comfort.
As expected, applying a naturalistic projection to $\autotraj{}$
results in a trajectory that curves along the outside of the road, replicating the behavior we see from human drivers.
In this case, both methods produce similar results upon projecting the original trajectory \hmzh{because each multimodal naturalistic subset captures a subset of the area of the unimodal subset, and the initial trajectory lies relatively close to the \nsets{} already.}

\subsection{Experiment 2: Intersection}
\label{ssec:exp2-intersection}


\cref{fig:inD-r22} captures a main road running through a four-way intersection and a pedestrian crossing, and the road is lined with parked cars.
We define $\dataset{}$ to include all moving vehicles beginning and ending in the eastbound lane.
Filtering based on the criteria in \cref{eq:exp-heuristic} results in 46 trajectories.
We will show that the \nset{} reflects how vehicles wait for crossing pedestrians and maintain distance from parked cars, without explicitly modeling those preferences.

\subsubsection{Configuration}
As with the first experiment, we choose to use \kmeans{}-constrained clustering and ensure that at most one cluster is invalid at any time step.
To select a value for $k$, we perform an analysis of the underlying naturalistic trajectory data in \cref{fig:inD-r22-trajectory-plots}, which plots the along-track distance of each trajectory over time.
A visual inspection of this data suggests two distances at which the plots become horizontal, indicating that trajectories tend to stop or slow at these places.
These regions map onto two locations in the lane: one is before the intersection, where vehicles wait for other vehicles to turn, and the other is a pedestrian crosswalk after the intersection.
As these stoppage points break the the lane into three regions, we select $k=3$.
We show the multimodal and unimodal naturalistic sets generated from $\dataset{}$ in \cref{fig:inD-r22-clusters2d-f245} and \cref{fig:inD-r22-hull2d-f245}.


\subsubsection{Analysis}
In this scenario, we can compare the ability of unimodal and multimodal naturalistic set generation to highlight complex behaviors without explicit modeling.
Applying unimodal set generation \cite{khan2024actnatural} to this scenario produces large and expanding naturalistic subsets $\{\wptset{t}\}$ which grow to span the entire lane due to the variety of speeds at which vehicles proceed along this lane (\cref{fig:inD-r22-hull2d-f245}).
A few outliers in the data (which can be seen in \cref{fig:inD-r22-trajectory-plots}) stop for a longer time and stretch the naturalistic subset $\wptset{t}$, which computes a convex hull over all points.
By contrast, multimodal naturalistic set generation produces a \nset{} with naturalistic subsets $\wptset{t}$ containing three convex sets because $k=3$ (\cref{fig:inD-r22-clusters2d-f245}).
In the beginning, before any trajectories pass through the intersection, these convex sets are clustered close together.
However, as vehicles pass through the intersection, larger gaps emerge between the convex sets near the points at which trajectories stop for oncoming vehicles and at the crosswalk.
Thus, our method generates \nsets{} and exposes multimodal behavior occurring in the intersection that was hidden within the unimodal method. 


Both methods account for the natural distance that moving vehicles have from parked cars.
Near the beginning of the task, both sets of \nset{} polygons are in the left portion of the lane, not centered.
More dramatically, as shown in the inset of \cref{fig:inD-r22-hull2d-f245} near the end of the task, the \nset{} is longer along the center of the lane than along the right side.
This observation, which also occurs in \cref{fig:inD-r22-clusters2d-f245}, indicates that vehicles move faster towards the center to avoid the parked red car.
Our method highlights the naturalistic preference to maintain a larger distance from parked vehicles, without explicitly modeling it.
Multimodal \nset{} generation produces tighter bounds because it finds similar points, thus eliminating space that would otherwise be included in a convex hull by connecting distant or dissimilar points.
In this case, we also see that the gaps in the convex sets at the points where vehicles stop also reduce the cumulative area of the convex sets from 62.00 m${}^2$ to 28.54 m${}^2$.
While this decrease does not always indicate a ``better'' \nset{}, in this case, it is beneficial to remove area from locations that are not occupied in human driving data.

Our new method highlights information about this intersection which exposes differences in the trajectories of vehicles which continue unencumbered, which pause, and which stop for longer. 
As such, our approach tightens the size of the \nset{} compared to previous approaches.

\begin{figure*}
\centering
\begin{minipage}{0.43\textwidth}
    \centering
    \includegraphics[width=0.7\textwidth]{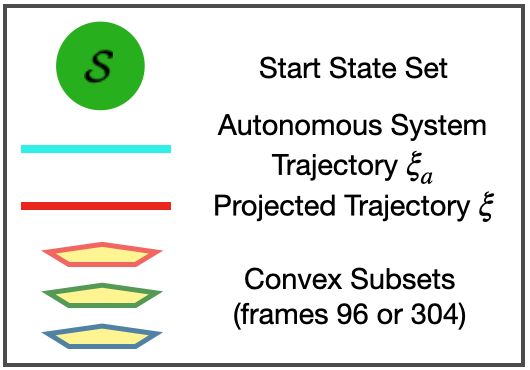}
    \subcaption{A legend for rounD, recording 0 (frames 96 and 304).}
    \label{fig:round-r0-legend}
\end{minipage}%
\hspace{5pt}
\begin{minipage}{0.12\textwidth}
    \centering
    \includegraphics[height=13em]{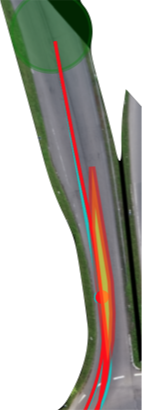}
    \subcaption{unimodal \cite{khan2024actnatural}}
    \label{fig:round-r0-hull2d-f96}
\end{minipage}%
\hspace{5pt}
\begin{minipage}{0.12\textwidth}
    \centering
    \includegraphics[height=13em]{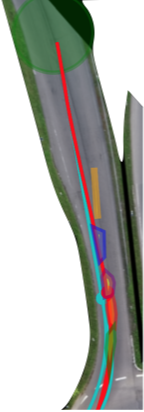}
    \subcaption{5-means.}
    \label{fig:round-r0-5-means-f96}
\end{minipage}%
\hspace{5pt}
\begin{minipage}{0.12\textwidth}
    \centering
    \includegraphics[height=13em]{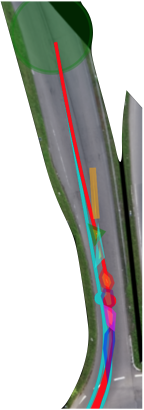}
    \subcaption{7-means}
    \label{fig:round-r0-7-means-f96}
\end{minipage}%
\hspace{5pt}
\begin{minipage}{0.12\textwidth}
    \centering
    \includegraphics[height=13em]{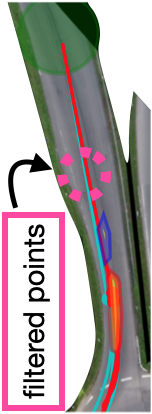}
    \subcaption{\texttt{HDBSCAN}.}
    \label{fig:round-r0-hdbscan-m3-e1-f96}
\end{minipage}

\begin{multicols}{2}
    \begin{subfigure}[t]{\columnwidth}
        \centering
        \includegraphics[width=0.95\textwidth]{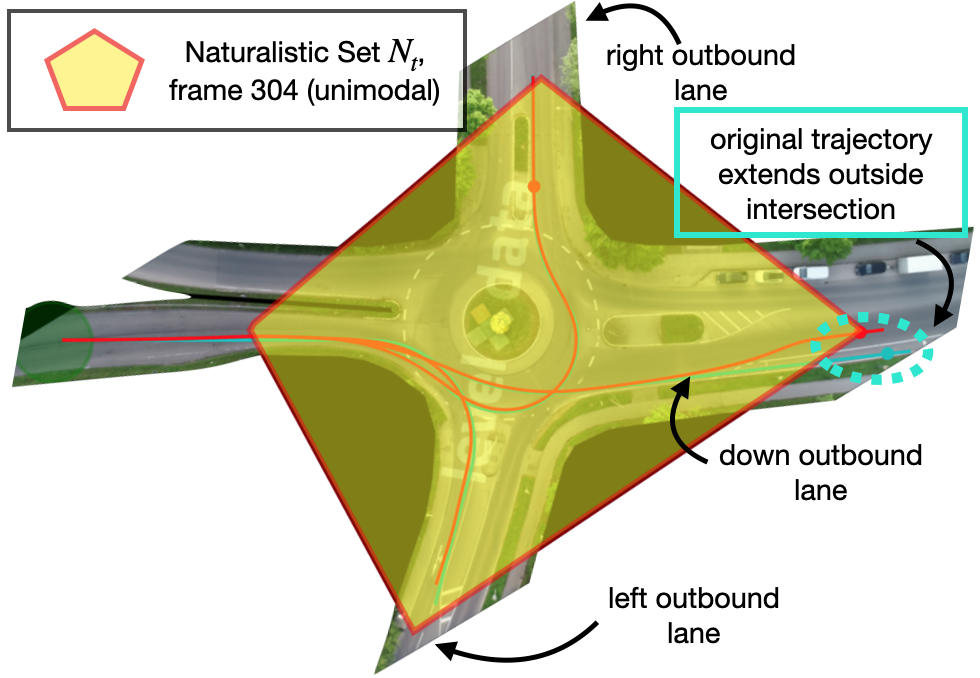}        
        \subcaption{Frame 304: Trajectories projected into $\natset{}{}$ using unimodal projection \citep{khan2024actnatural}.}
        \label{fig:round-r0-hull2d-f304}
    \end{subfigure}

    \begin{subfigure}[t]{\columnwidth}
        \centering
        \includegraphics[width=0.95\textwidth]{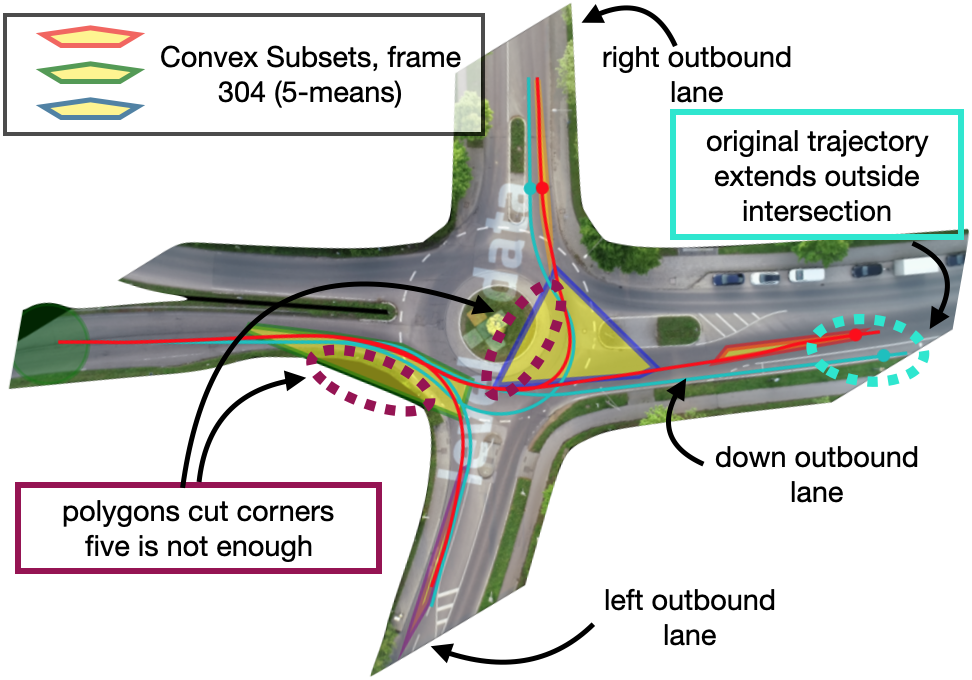}
        \subcaption{Frame 304: Autonomous trajectories projected into $\natset{}{}$ made with 5-means clusters.}
        \label{fig:round-r0-5-means-f304}
    \end{subfigure}

\end{multicols}

\begin{multicols}{2}

    \begin{subfigure}[t]{\columnwidth}
        \centering
        \includegraphics[width=0.95\textwidth]{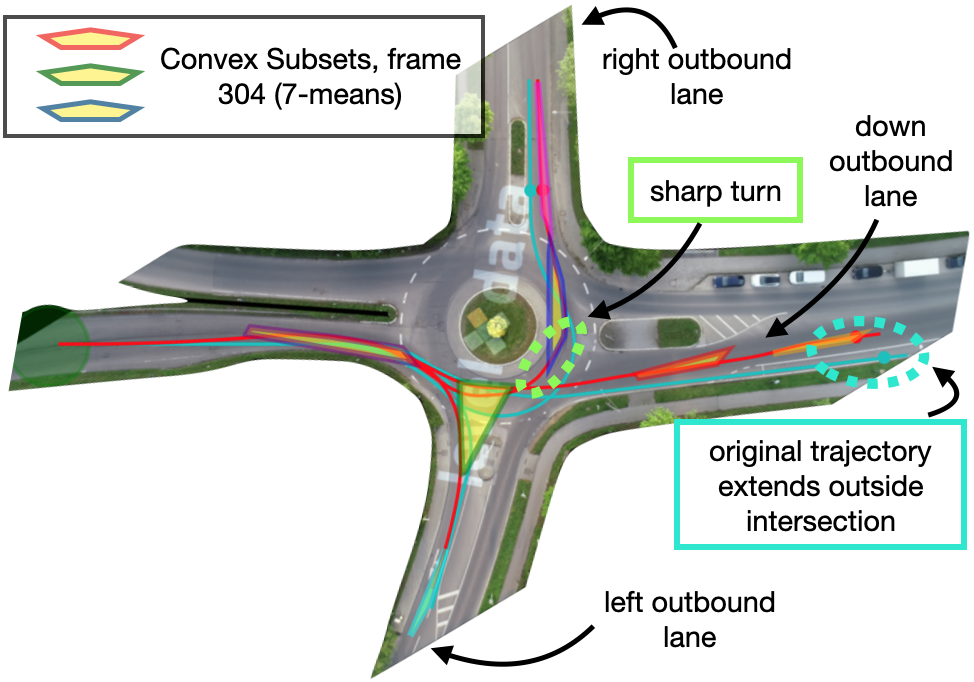}
        \subcaption{Frame 304: Autonomous trajectories projected into $\natset{}{}$ made with 7-means clustering.}
        \label{fig:round-r0-7-means-f304}
    \end{subfigure}

    \begin{subfigure}[t]{\columnwidth}
        \centering
        \includegraphics[width=0.95\textwidth]{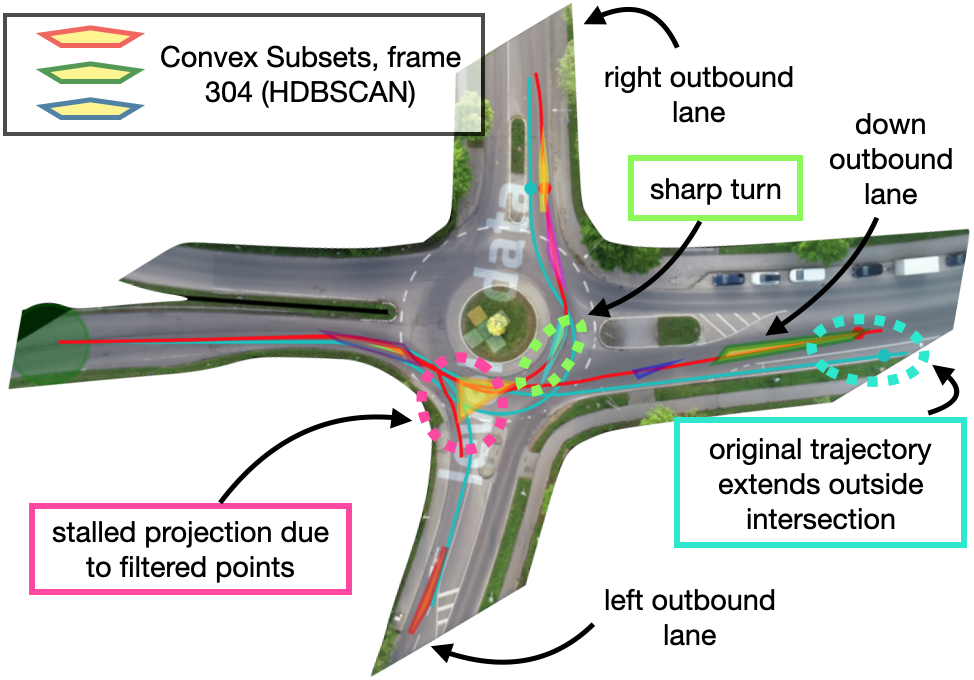}
        \subcaption{Frame 304: Autonomous trajectories projected into $\natset{}{}$ made with \texttt{HDBSCAN} clustering.
        \label{fig:round-r0-hdbscan-m3-e1-f304}}
    \end{subfigure}
\end{multicols}

\caption{
A comparison on the rounD dataset (recording 0) between the performance of \citep{khan2024actnatural} and our algorithm configured to use different clustering algorithms (5-means, 7-means, and \texttt{HDBSCAN}).
Three trajectories (shown in cyan, and referred to as left, down, and right)  which start in the same inbound lane but exit from different outbound lanes are projected (shown in red) into \nsets{} generated with four naturalistic set generation methods (unimodal, 5-means, 7-means, \texttt{HDBSCAN}).
\cref{fig:round-r0-hull2d-f96,fig:round-r0-5-means-f96,fig:round-r0-7-means-f96,fig:round-r0-hdbscan-m3-e1-f96} 
show the naturalistic subset ($\wptset{96}$) early in the recording at frame 96, and
\cref{fig:round-r0-hull2d-f304,fig:round-r0-5-means-f304,fig:round-r0-7-means-f304,fig:round-r0-hdbscan-m3-e1-f304} show $\wptset{304}$ later in the recording (frame 304).
We highlight parts of the \nsets{} or trajectories that demonstrate how well each method works.
}
\label{fig:exp-3-round-r0}
\end{figure*}

\subsection{Experiment 3: Roundabout}
\label{ssec:exp3-roundabout}

\cref{fig:exp-3-round-r0} captures a roundabout intersection with four two-way single-lane roads entering and exiting the roundabout towards each cardinal direction. 
We define $\dataset{}$ to include all inbound moving vehicles beginning in the north inbound lane, where $\startset{}$ is given by the green circle and we do not specify $\finalset{}$.
Thus, $\dataset{}$ includes 63 vehicles proceeding counterclockwise around the roundabout, exiting in one of the 
leftward, downward, or rightward directions.
We show that the \nset{} reflects behavior introduced by the branching paths; a vehicle may choose to take the leftward exit, or proceed along the roundabout, at which point similar decisions exist for the downward and rightward directions.
Example hand-crafted trajectories which exit in each direction are projected into the multimodal \nset{}.
Given the various paths available to drivers in this scenario, we find that a multimodal representation is critical for modeling the behavior of vehicles in this intersection.

\subsubsection{Configuration}
In this experiment, we test four different multimodal \nset{} generation algorithms.
Each uses a different clustering algorithm: unimodal naturalistic set generation \cite{khan2024actnatural} (which is equivalent to using 1-means clustering), 5-means-constrained clustering, 7-means-constrained clustering, and \texttt{HDBSCAN}.
We configure these clustering methods as described in \cref{ssec:exp-nset-generation}.

\subsubsection{Analysis}

\begin{figure*}
\centering
\begin{multicols}{3}

    \begin{subfigure}[t]{\columnwidth}
        \centering
        \includegraphics[width=\columnwidth,height=12em]{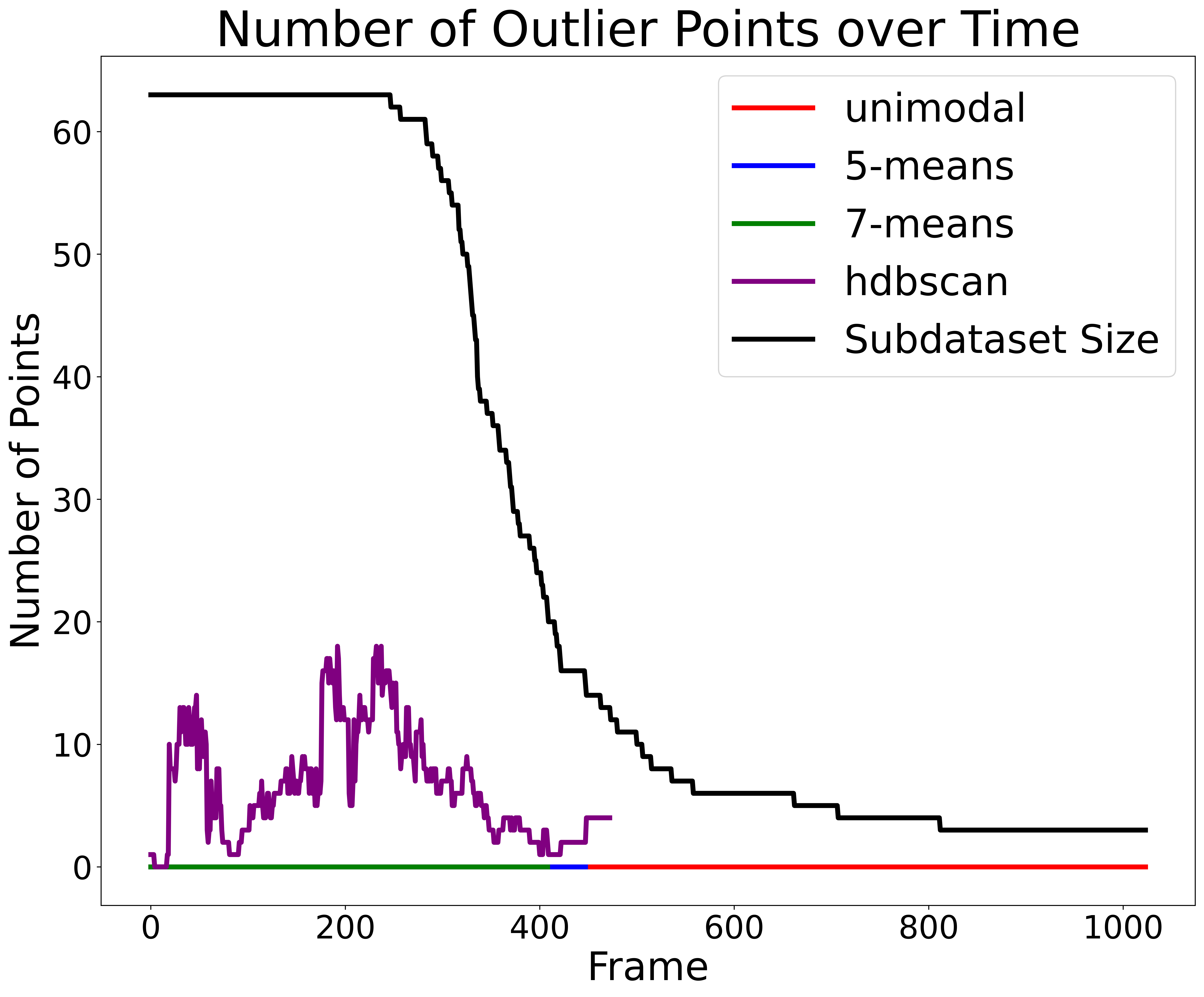}
        \subcaption{Num. total/invalid points used to generate $\wptset{t}$. \label{fig:metrics-num-points}}
        
    \end{subfigure}
    
    \begin{subfigure}[t]{\columnwidth}
        \centering
        \includegraphics[width=\columnwidth,height=12em]{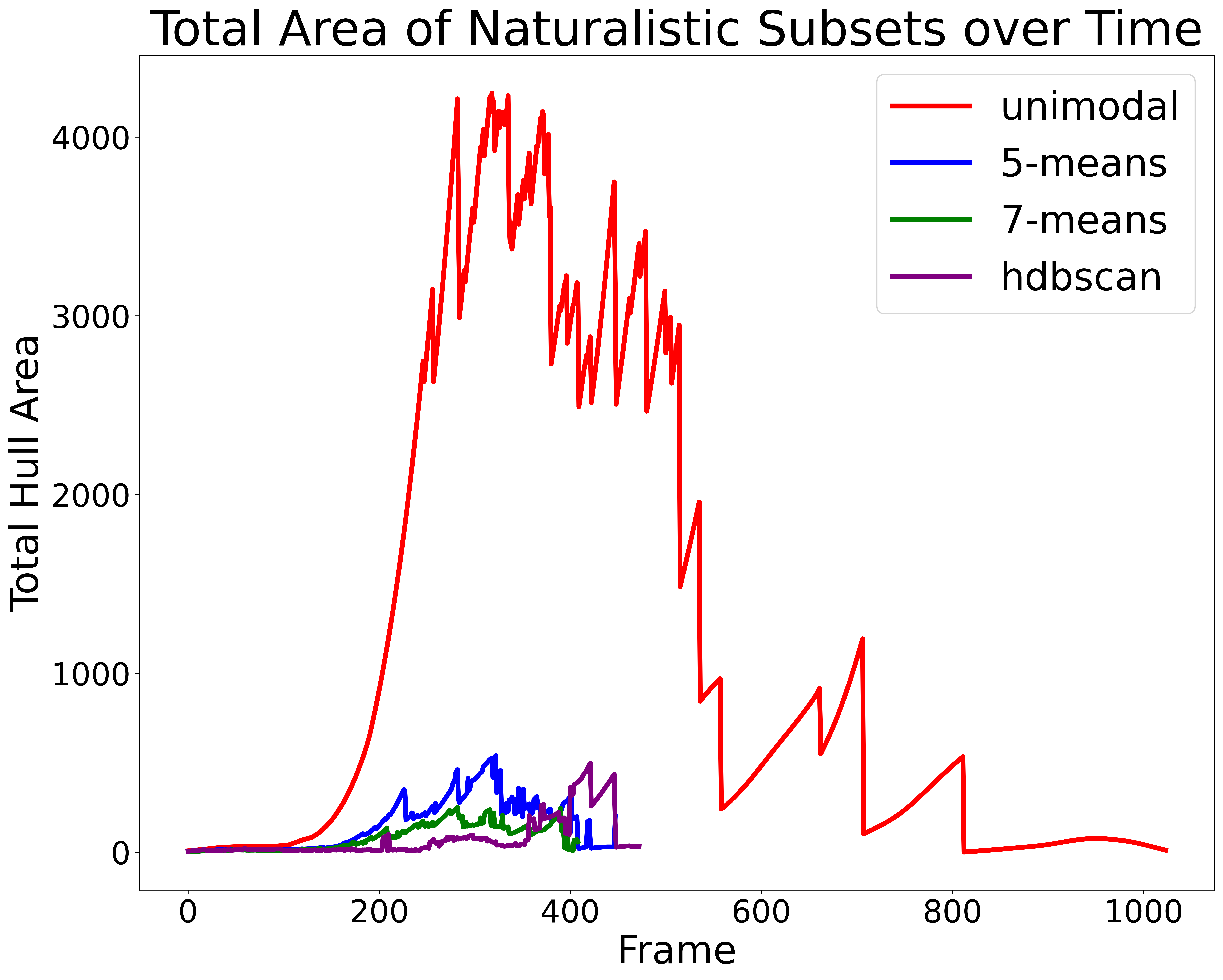}
        \subcaption{Total area of the naturalistic subsets. \label{fig:metrics-area}}
        
    \end{subfigure}

    \begin{subfigure}[t]{\columnwidth}
        \centering
        \includegraphics[width=\columnwidth,height=12em]{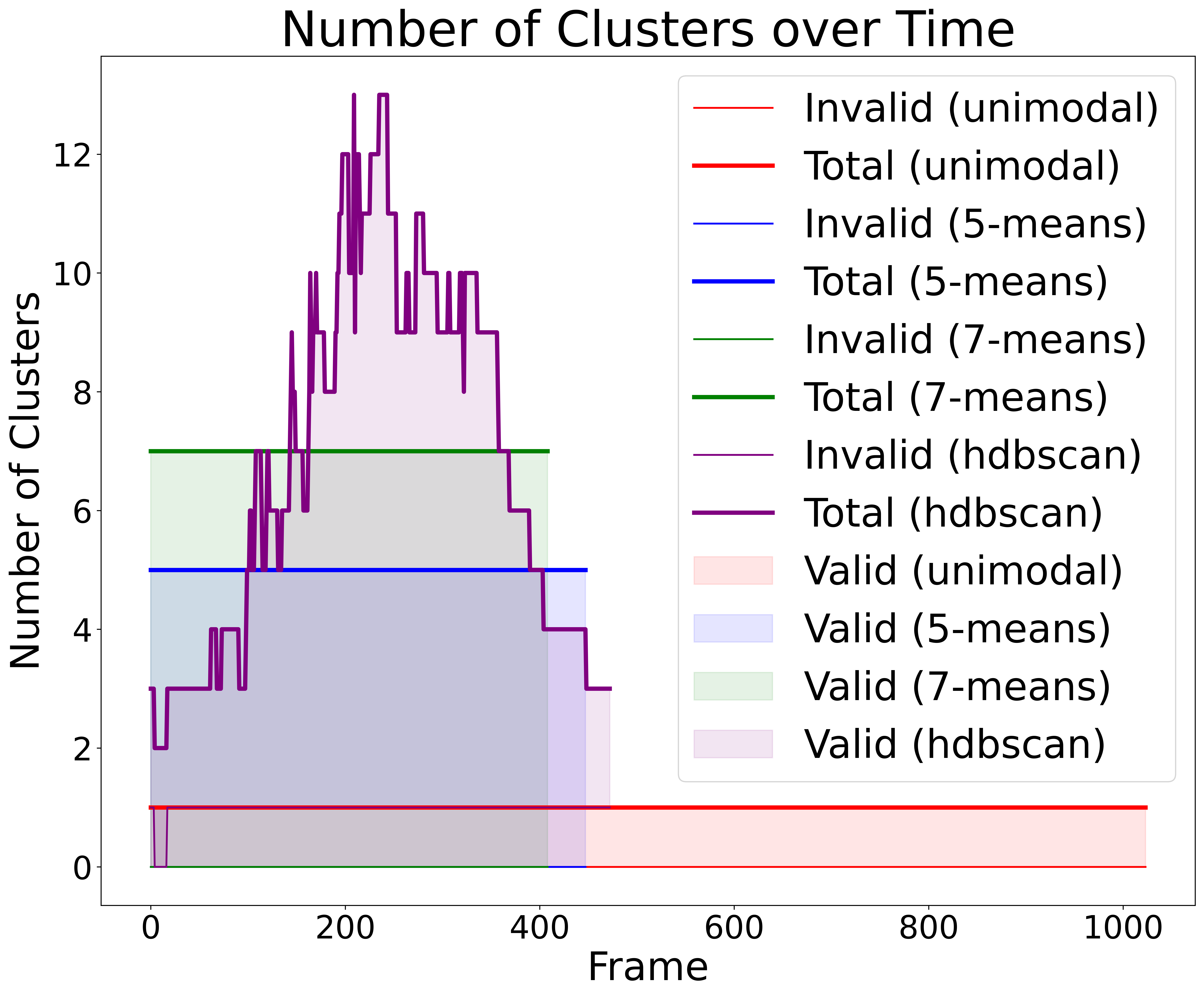}
        \subcaption{Num. clusters ($k_t$) for each clustering method.}
        \label{fig:metrics-num-clusters}
    \end{subfigure}
\end{multicols}
\vspace{-2em}
\caption{
A comparison on the rounD dataset (recording 0) between the performance of \citep{khan2024actnatural} and our algorithm configured to use different clustering algorithms (5-means, 7-means, and \texttt{HDBSCAN}).
\eqref{fig:metrics-num-points} shows the total number of points used to generate $\dataset{t}$ and the number of outlier points over time for each method.
Only \texttt{HDBSCAN} has outlier rejection, and these points all reside in the ``noisy'' set.
\eqref{fig:metrics-area} shows the sum of the area of each convex set within $\wptset{t}$.
\eqref{fig:metrics-num-clusters} shows the number of clusters, $k_t$, produced by each method over time, which is a (configured) constant value except for \texttt{HDBSCAN}, which infers it at each time.}
\vspace{-1em}
\label{fig:exp-3-nset-metrics}
\end{figure*}

In \cref{fig:exp-3-round-r0}, we show the \nsets{} generated for these four algorithms at frames, 96 and 304.
Between these two times, and the four naturalistic projection methods, we present a broader view of how multimodal naturalistic projection performs under different distributions of the human driving data.

\textbf{Multimodal \nsets{} represent this scenario more faithfully than unimodal naturalistic sets.}
In \cref{fig:round-r0-hull2d-f304}, the naturalistic subset $\wptset{304}$ encompasses the entire roundabout, nondrivable regions inside and outside the intersection, and lanes oriented in the wrong direction.
Clearly, a single convex hull is insufficient to describe the complex behavior here.
Additionally, as shown in \cref{fig:metrics-area}, unimodal set generation produces a polygon which has two orders of magnitude more area than the total area of the polygons produced by our multimodal techniques.
At each time $t$, \textbf{each multimodel method produces polygons we can associate with features of the intersection that vehicles drive on  and which follow the branching structure of trajectories traversing the lanes in the roundabout in \cref{fig:exp-3-round-r0}.}
Observe $\wptset{304}$ of \cref{fig:round-r0-5-means-f304,fig:round-r0-7-means-f304,fig:round-r0-hdbscan-m3-e1-f304}, for example, which contain polygons corresponding to the inbound lane, the outbound lanes, and points where vehicle trajectories diverge to continue around the roundabout or exit to an outbound lane.

\begin{figure*}
\centering
\begin{multicols}{3}
    \begin{subfigure}[t]{\columnwidth}
        \centering
        \includegraphics[width=\columnwidth,height=11em]{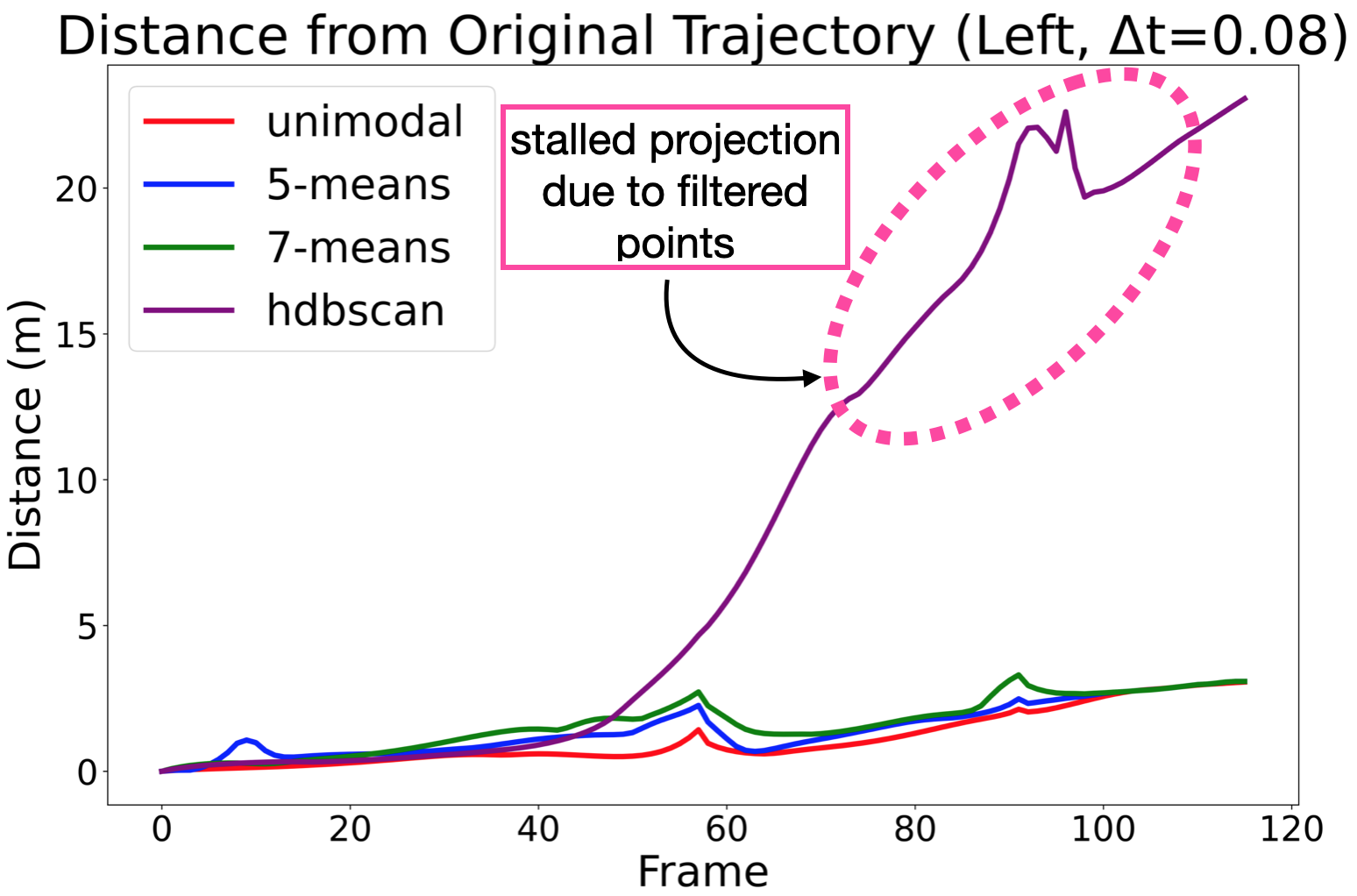}
        \subcaption{Dist. from original to projected left trajectory. \label{fig:round-r0-left-projected-dist}}
    \end{subfigure}
    
    \begin{subfigure}[t]{\columnwidth}
        \centering
        \includegraphics[width=\columnwidth,height=11em]{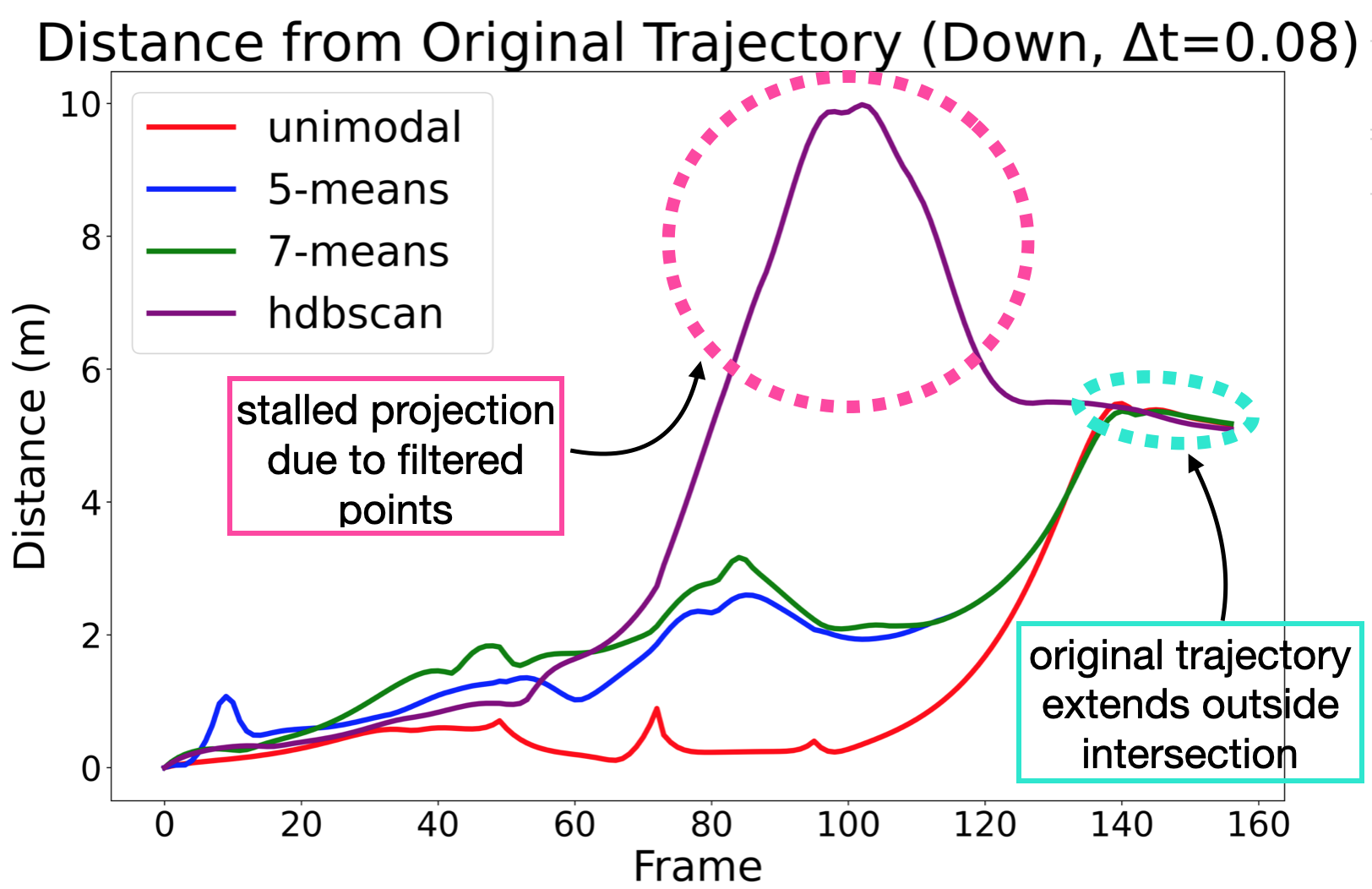}
        \subcaption{Dist. from original to projected down trajectory. \label{fig:round-r0-down-projected-dist}}
    \end{subfigure}

    \begin{subfigure}[t]{\columnwidth}
        \centering
        \includegraphics[width=\columnwidth,height=11em]{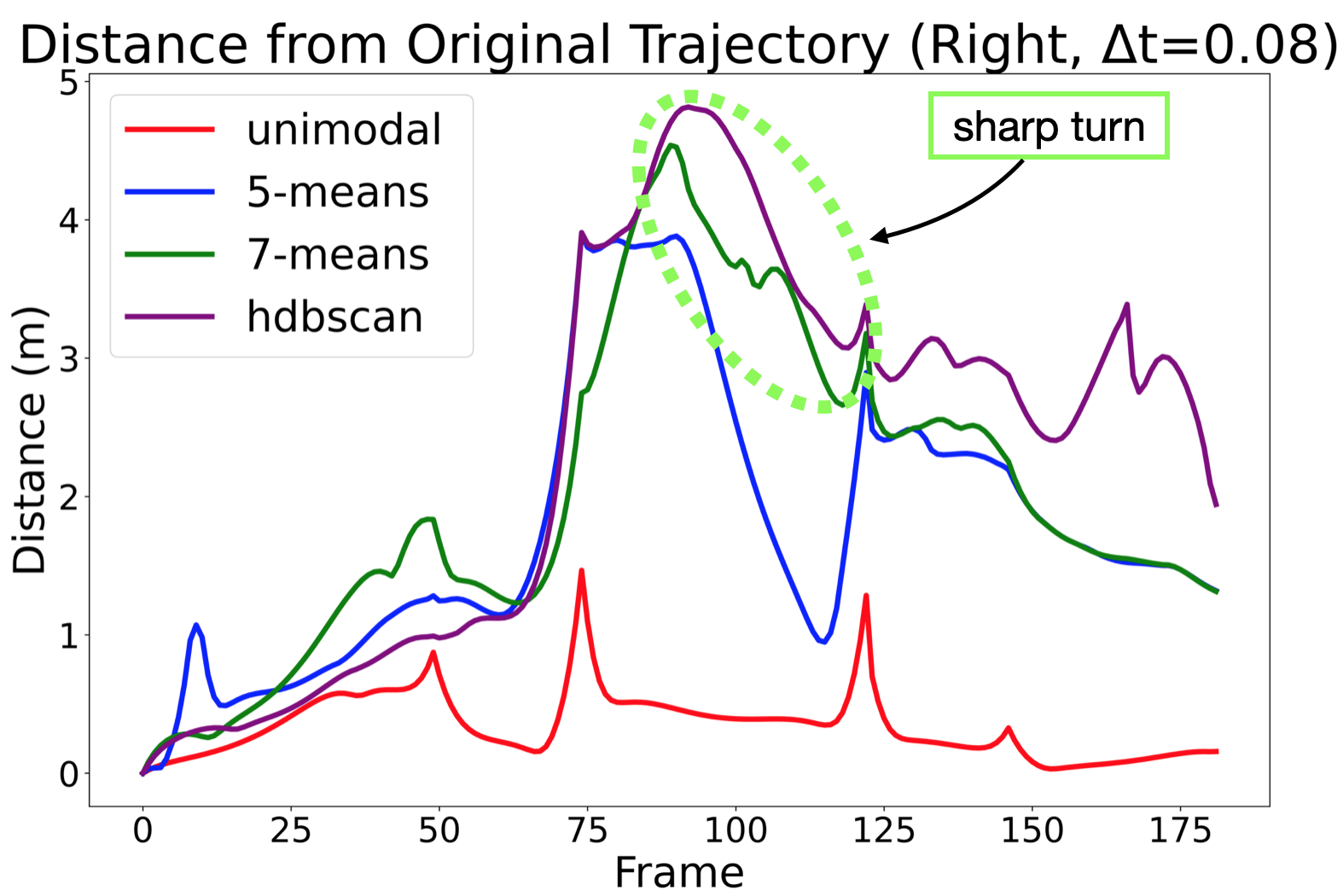}
        \subcaption{Dist. from original to projected right trajectory. \label{fig:round-r0-right-projected-dist}}
    \end{subfigure}
\end{multicols}
\vspace{-2em}
\caption{Distances between the states of each original trajectory $\autotraj{}$ and each projected, naturalistic trajectory $\traj{}$.
We highlight relevant phenonema related to the performance of each method.
\label{fig:round-r0-projected-distances}}
\vspace{-1em}
\end{figure*}

Early in the scenario for the inbound lane, the unimodal and \texttt{HDBSCAN} set generation methods produce fewer clusters (as shown in \cref{fig:metrics-num-clusters}), meaning that each cluster contains more points.
In \cref{fig:round-r0-hull2d-f96}, the unimodal approach produces a naturalistic subset $\wptset{96}$ containing one polygon that covers the whole inbound lane, and in \cref{fig:round-r0-hdbscan-m3-e1-f96}, the subset $\wptset{96}$ produced using \texttt{HDBSCAN} covers the lane with two polygons.
In contrast, 5-means (\cref{fig:round-r0-5-means-f96}) and 7-means (\cref{fig:round-r0-7-means-f96}) produce a subset $\wptset{96}$ with a fixed number of 5 and 7 clusters, respectively, resulting in fewer points per cluster.
In \cref{fig:metrics-num-clusters}, the number of non-noisy clusters produced by \texttt{HDBSCAN} fluctuates between 2-3 in the first 125 frames because \texttt{HDBSCAN} identifies the number of clusters automatically rather than fixing them.

\textbf{Increasing the number of clusters improves the tightness of the convex sets where vehicles' routes diverge.}
For example, consider the middle cluster of \cref{fig:round-r0-5-means-f304}. This large cluster cuts through a signficant portion of the roundabout circle, which vehicles never drive through, and therefore can not be naturalistic.
Similarly, as the vehicles exit the inbound lane (circled in purple in \cref{fig:round-r0-5-means-f304}), we see that polygons in that area get very close to the lane edges due to the curvature of the road.
This behavior indicates that we may not have enough polygons to represent the behavior in the intersection.
By contrast, in \cref{fig:round-r0-7-means-f304}, setting $k=7$ instead of $k=5$
helps avoid overapproximations, as more clusters enable splitting the points in a way that reduces cutting across corners of the intersection.

\subsubsection{Projection}
\label{sssec:exp-projection}

Each subfigure in \cref{fig:exp-3-round-r0} depicts three trajectories that travel from the start set to exit in each outbound lane (shown in cyan).
Each subfigure further shows an associated trajectory (in red) projected using our method into the multimodal \nset{} depicted in the subfigure.
We make a few observations about the resulting projections.

First, due to the large (and clearly not naturalistic) \nset{} produced by unimodal set generation in \cref{fig:round-r0-hull2d-f304}, the original trajectory is largely within $\natset{}{}$.
As affirmed by \cref{fig:round-r0-projected-distances}, unimodal projection (in red) consistently produces a trajectory with the lowest distance from the original trajectory.
The distance is generally less than 1 meter, except where the original trajectory extends outside the intersection (circled in cyan) as shown most clearly in \cref{fig:round-r0-down-projected-dist} and where we see spikes indicating a jump in the convex hull as one of the underlying data points exits the roundabout.

Projecting into \nsets{} generated by 5-means-constrained and 7-means-constrained produce projection results which differ similarly in distance from the original.
However, we note a significant difference in distance in the right trajectory, highlighted in light green in \cref{fig:round-r0-right-projected-dist}.
Here, we see that increasing the number of clusters results in a sharper turn going around the roundabout for the projection enforcing naturalistic behavior generated using 7-means-constrained (highlighted in light green in \cref{fig:round-r0-7-means-f304}).
Thus, \textbf{we note a tradeoff between projections into \nsets{} with fewer clusters (with more points per cluster) sometimes including unoccupied space within $\natset{}{}$ and those with more clusters (and thus fewer points per cluster) resulting in a less smooth projection.}

We see this behavior more drastically in the \nsets{} generated with \texttt{HDBSCAN}, which infers between 2 and 13 clusters (\cref{fig:metrics-num-clusters}).
The effect is visible in the projections of the leftward and downward trajectories, where the distances between them and the original trajectories become significantly higher than those of the other methods (circled in pink in \cref{fig:round-r0-left-projected-dist}, \cref{fig:round-r0-down-projected-dist}).
The corresponding projections (also highlighted in pink) in \cref{fig:round-r0-hdbscan-m3-e1-f304} introduce awkward sharpness in the trajectories that cause them to stall.
\hmzh{The choice of $\gamma$, which influences the control cost, may also contribute to this lack of movement.}

Moreover, \texttt{HDBSCAN} also filters out ``noisy'' points (highlighted in pink in \cref{fig:round-r0-hdbscan-m3-e1-f96,fig:round-r0-hdbscan-m3-e1-f304}).
While the capability of filtering outliers is important, in this case, its effect results in many fewer points per cluster---at times, \texttt{HDBSCAN} labels over a quarter of the points as noisy and thus invalid (\cref{fig:metrics-num-points}), exacerbating the issue of having too many clusters for the amount of data.

Overall, we note that our method produces trajectories that satisfy the naturalistic constraints and modify behavior to appear more naturalistic.
However, there are tradeoffs in choosing the number of clusters and how aggressively noisy points are filtered out.
Fewer clusters can result in ``cutting corners'' and including regions that humans never occupy, whereas more clusters and excessive filtering can result in fewer points per cluster and thus sharper trajectories.

\subsection{Projection Runtime and Quality}
\label{ssec:exp-timing}

\begin{table*}[!ht]
\centering
\resizebox{0.9\textwidth}{!}{
\begin{minipage}{\textwidth}
\centering
\begin{tabular}{|c|c|c|c|c|c|}
\hline
\textbf{Clustering Method} & \textbf{Downsample} & \textbf{Frame Skip} & \textbf{Left Traj.} & \textbf{Down Traj.} & \textbf{Right Traj.} \\
 & \textbf{Rate} & \textbf{Rate ($\Delta \phi$)} & \textbf{Runtime (s)} & \textbf{Runtime (s)} & \textbf{Runtime (s)}\\
\hline\hline
Unimodal \cite{khan2024actnatural} & 2 & 1 & 0.0471 & 0.0305 & 0.0340 \\ \hline
Unimodal \cite{khan2024actnatural} & 2 & 2 & 0.0437 & 0.0264 & 0.0268 \\ \hline
Unimodal \cite{khan2024actnatural} & 2 & 4 & 0.0424 & 0.0212 & 0.0213 \\ \hline
Unimodal \cite{khan2024actnatural} & 2 & 8 & 0.0811 & 0.0187 & 0.0220 \\ \hline\hline
5-means-constrained & 2 & 1 & 36.5380 & 41.9413 & 35.6198 \\ \hline
5-means-constrained & 2 & 2 & 4.0508 & 10.1675 & 14.7685 \\ \hline
5-means-constrained & 2 & 4 & 0.9592 & 1.3842 & 1.5393 \\ \hline
5-means-constrained & 2 & 8 & 0.4524 & 0.6197 & 0.8781 \\ \hline\hline
7-means-constrained & 2 & 1 & 29.0983 & 56.2113 & 135.7752 \\ \hline
7-means-constrained & 2 & 2 & 13.7343 & 13.6035 & 19.1819 \\ \hline
7-means-constrained & 2 & 4 & 1.5443 & 3.4226 & 6.2815 \\ \hline
7-means-constrained & 2 & 8 & 0.3392 & 0.6096 & 1.2816 \\ \hline\hline
\texttt{HDBSCAN} & 2 & 1 & 90.2838 & 28.2979 & 56.3699 \\ \hline
\texttt{HDBSCAN} & 2 & 2 & 18.2206 & 8.5388 & 11.3902 \\ \hline
\texttt{HDBSCAN} & 2 & 4 & 0.9182 & 1.4176 & 2.5253 \\ \hline
\texttt{HDBSCAN} & 2 & 8 & 0.2468 & 0.8348 & 0.6570 \\ \hline
\end{tabular}
\end{minipage}
}
\vspace{1em}
\label{tab:timing_data}
\caption{Computation Time for Mixed-Integer Projection of Multiple Trajectories (rounD, Recording 0)\vspace{-2em}}
\end{table*}

\begin{figure*}
\centering
\begin{multicols}{3}
    \begin{subfigure}[t]{\columnwidth}
        \centering
        \includegraphics[width=0.97\columnwidth, height=10em]{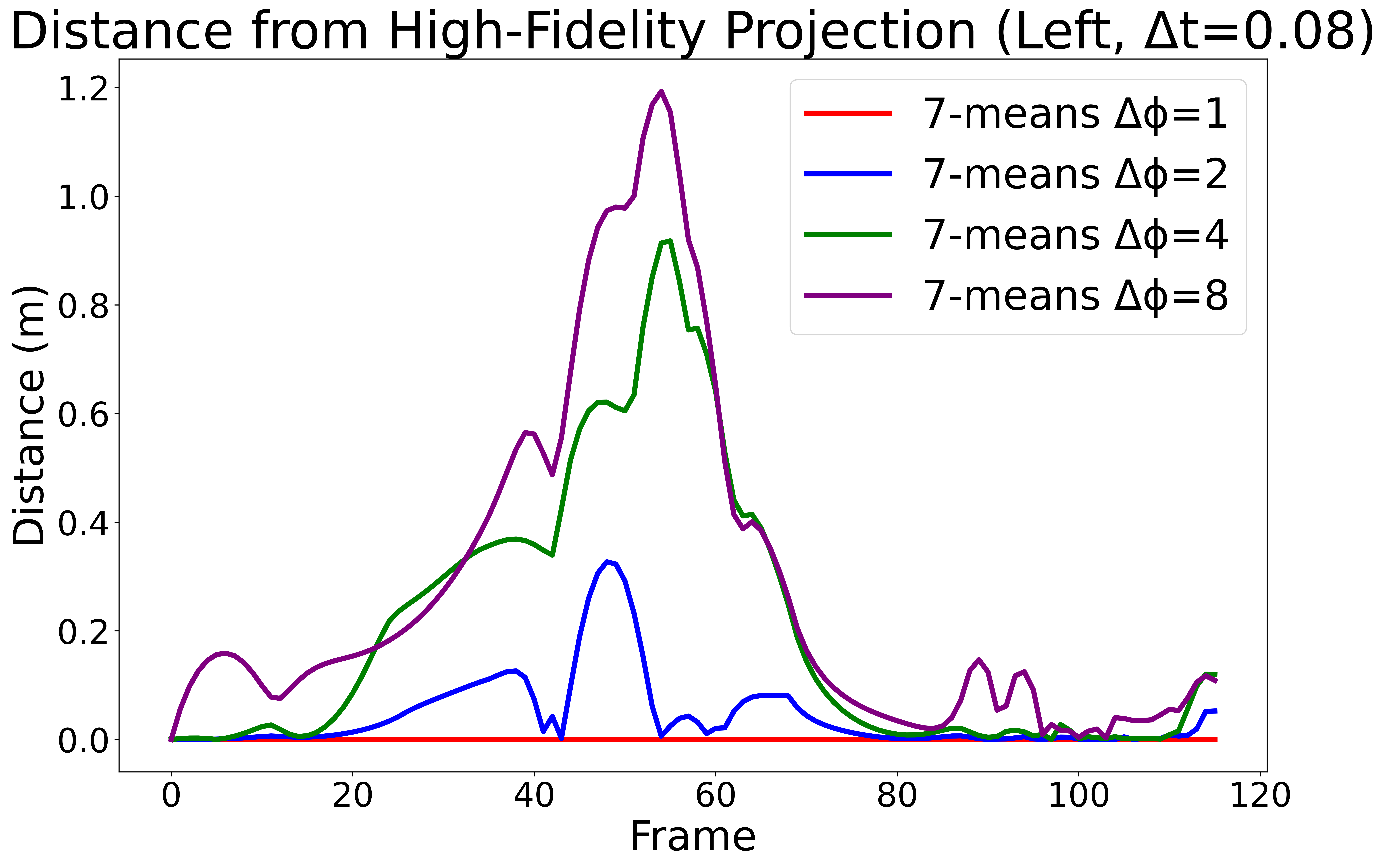}
        \subcaption{Left traj. projection quality with varying $\Delta \phi$.}
        \label{fig:round-r0-left-proj-quality}
    \end{subfigure}
    
    \begin{subfigure}[t]{\columnwidth}
        \centering
        \includegraphics[width=\columnwidth]{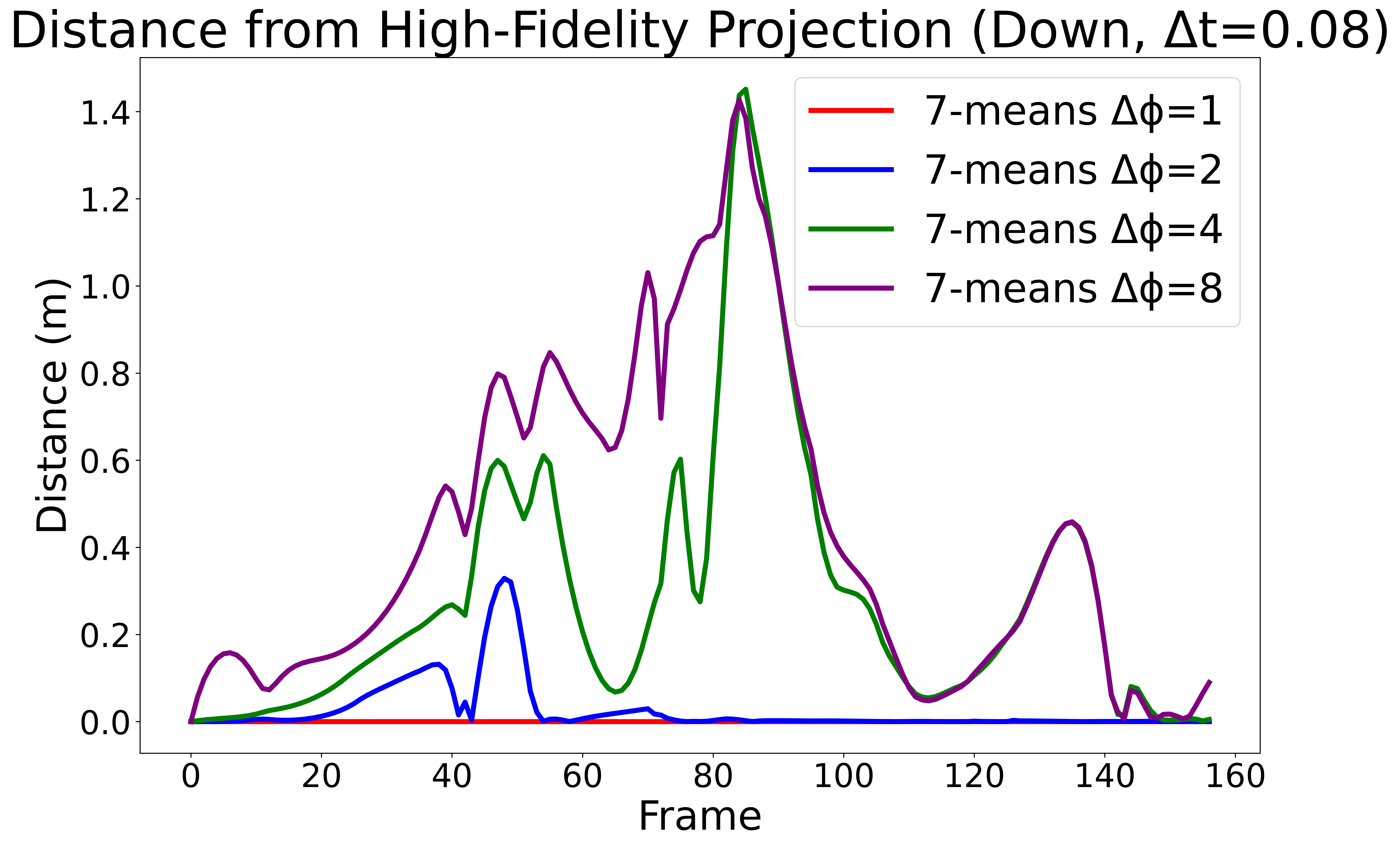}
        \subcaption{Down traj. projection quality with varying $\Delta \phi$.}
        \label{fig:round-r0-down-proj-quality}
    \end{subfigure}
    
    \begin{subfigure}[t]{\columnwidth}
        \centering
        \includegraphics[width=\columnwidth, height=10em]{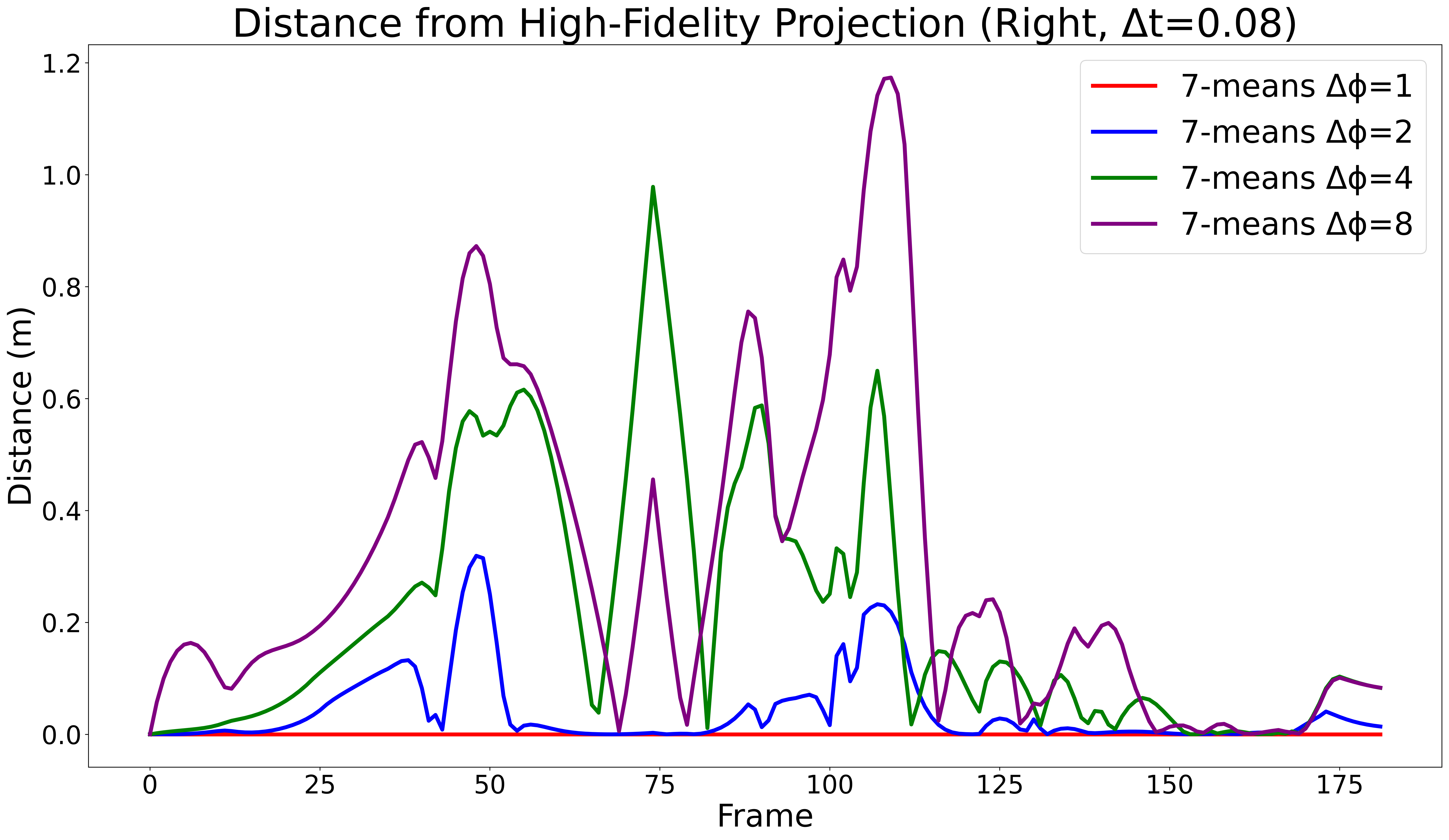}
        \subcaption{Right traj. projection quality with varying $\Delta \phi$.}
        \label{fig:round-r0-right-proj-quality}
    \end{subfigure}
\end{multicols}
\vspace{-1.5em}
\caption{Results using 7-means-constrained naturalistic set generation to test projection quality as projection fidelity decreases.
The results depict the relative distances between a high-fidelity projection (where $\Delta \phi = 1$) for which the naturalistic constraint is enforced at every frame and lower-fidelity projections where it is enforced at a lower rate, i.e. $\Delta \phi = 2, 4, 8$. \vspace{-2em}
\label{fig:round-r0-proj-quality}}
\end{figure*}


In this section, we report the computational runtime of the mixed-integer quadratic problem used in our projection technique.
Solving MIQPs is generally NP-hard, though highly optimized solvers exist \cite{gurobi} that use techniques like branch and bound \cite{land2010branchbound} to quickly solve these problems.
Nevertheless, as runtimes can still vary, we provide a method of trading off runtime against solution quality.
Specifically, the most time-consuming constraints for a solver to enforce are dynamics constraints \cref{eq:projection-opt-dynamics-exp} and the naturalistic set constraints  in \cref{eq:at-least-1-exp} and \cref{eq:projection-opt-nset-exp}.
The inD and rounD datasets provide data sampled at \SI{25}{\hertz} (\SI{0.04}{\second} period), which we downsample to reduce the number of constraints that must be enforced without significantly reducing solution quality.
To enable further tuning, we provide a parameter $\Delta \phi$ that reduces the rate of enforcement of the naturalistic behavior constraints. Thus, rather than enforce \cref{eq:at-least-1-exp} and \cref{eq:projection-opt-nset-exp} for all times $t \in [T+1]$, we enforce them for all times $t \in \{\Delta \phi, 2\Delta \phi, \ldots, \horizon{}\}$.

We report the results of this tuning in \cref{tab:timing_data} for the three trajectories we project in \cref{fig:exp-3-round-r0}.
In particular, enforcing naturalistic constraints half the time can result in up to an 18x speedup (with clusters from \texttt{HDBSCAN}) in projection time, and many of the results drop below 1 second and are thus useful in many real-time applications.
Moreover, in \cref{fig:round-r0-proj-quality}, we report the changes in the solution quality against the highest fidelity projection, which enforces all constraints at \SI{12.5}{\hertz}.
For this particular method, we find that the difference between the high- and lower-fidelity projected trajectories typically increases with $\Delta \phi$. In this example, an eight-fold increase results in at most 1.4 meters of distance.

\section{Discussion and Limitations}
\label{sec:limitations}

We present our approach as a practical next step that extends unimodal naturalistic projection \cite{khan2024actnatural} to be
more capable of representing complex behaviors with \nsets{} and projecting trajectories into them.
Specifically, we introduce a representation by which multimodality and nonconvexity can be modeled, and by which autonomous trajectories can be projected into it.
Our mechanism for doing so involves representing complex behavior as a union of convex sets, and we compute each convex set based on the results of a clustering algorithm applied to a naturalistic dataset.
Our method identifies naturalistic behavior without explicit modeling, and reproduces it through projection into the \nset{}.
In this section, we discuss some limitations of our approach, and future research directions that may address them.



\noindent\textbf{Interaction.} As previously discussed, our approach identifies \hmzh{human tendencies to yield in certain locations when interacting with other agents.}
However, as with \cite{khan2024actnatural}, it does not directly account for interaction.
In practice, a more direct understanding of how human drivers interact with one another is critical for applying this approach in real-world applications.
This work's extension to multimodal behaviors and nonconvex data is critical for modeling (highly nonconvex) interactive behavior.
Thus, we anticipate that future work \hmzh{could address this weakness by analyzing interaction using naturalistic sets computed jointly on the data of the interacting agents.}



\noindent\textbf{Undesirable Road Behaviors.} \hmzh{Although our method flexibly replicates naturalistic behavior, it raises an important question: should all forms of naturalistic behavior be reproduced? 
Take, for instance, a scenario in which road users routinely ignore stop signs. 
In such cases, it is more desirable for an autonomous planner to prioritize adherence to legal standards rather than reinforcing unlawful behavior.
Our approach is versatile enough to incorporate these additional preferences by introducing constraints into \cref{eq:projection-opt} and \cref{eq:projection-opt-exp}.}


\noindent\textbf{Real-time Data Collection and Usage.}
In this work, we propose that data collection, filtration, and naturalistic set generation occur offline, while projection occurs online.
For example, data on how vehicles drive in a certain intersection could be collected by a camera with a high vantage point, or by multiple cars in a fleet which drive through the intersection.
This data could then be aggregated and used to generate a \nset{} which could reflect a set period of history which captures on-road conditions in the intersection, i.e. the last four hours of driving behavior.
As with \cite{khan2024actnatural}, vehicles using naturalistic projection would then be able to project their own autonomous trajectories into the \nset{}.
Future work is needed to experimentally validate this proposed method of operation \emph{in situ}.

\noindent\textbf{Outliers.}
While outliers are known to inflate convex hulls, methods exist to compute convex hulls while remaining robust to noise in the underlying data \citep{fischler1981ransac}.
Some of these methods are built-in to specific clustering algorithms like \texttt{HDBSCAN} \citep{campello2015hierarchical}.
Nevertheless, when used aggressively as in \cref{fig:round-r0-hdbscan-m3-e1-f304}, these methods can result in awkward projections due to the low quantity of points in some clusters.
Future work in this direction must seek out how to best balance these concerns.

\section*{Acknowledgment} We thank Adam Thorpe for his ideas and thoughts on technical directions that led to this work, Cade Armstrong and Ryan Park for support in generating trajectories.
\printbibliography  

\vspace{-5em}
\begin{IEEEbiography}[{\includegraphics[width=1in,height=1.25in,clip,keepaspectratio]{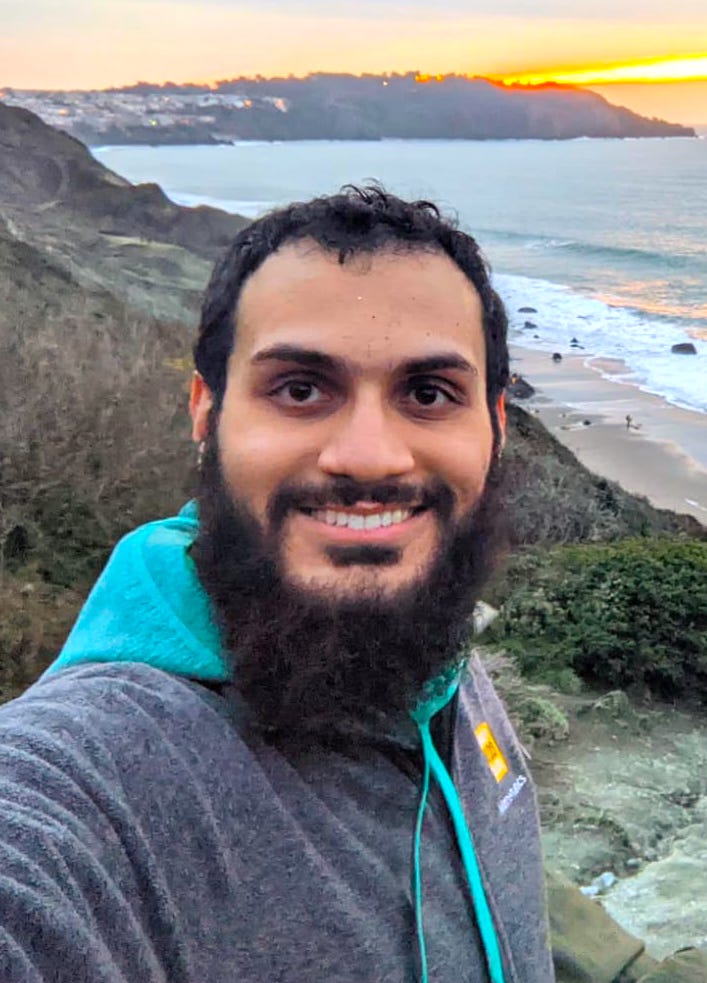}}]{Hamzah I. Khan} (Student Member, IEEE) received a B.S. degree in robotics from Harvey Mudd College, and a Master's degree from the University of Texas in Austin.
He is a doctoral candidate with Prof. David Fridovich-Keil in the CLeAR Lab at UT Austin, studying interactive and strategic decision-making, planning, and control.
\end{IEEEbiography}
\vspace{-5em}
\begin{IEEEbiography}[{\includegraphics[width=1in,height=1.25in,clip,keepaspectratio]{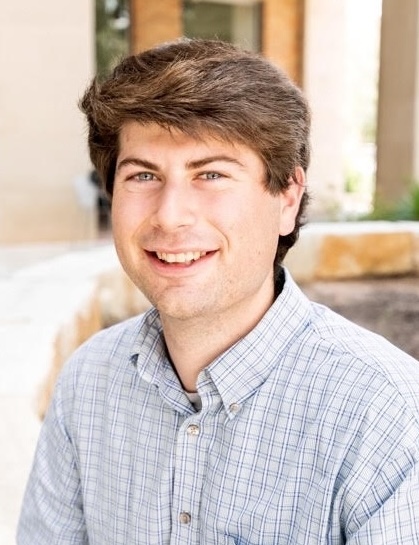}}]{David Fridovich-Keil} (Member, IEEE) received the B.S.E. degree in electrical engineering from Princeton University, and the Ph.D. Degree from the University of California, Berkeley. He is an Assistant Professor in the Department of Aerospace Engineering and Engineering Mechanics at the University of Texas at Austin. Fridovich-Keil is the recipient of an NSF Graduate Research Fellowship and an NSF CAREER Award.
\end{IEEEbiography}




\end{document}